\documentclass[lettersize,journal]{IEEEtran}
\usepackage{amsmath,amsfonts}
\usepackage{array}
\usepackage[caption=false,font=normalsize,labelfont=sf,textfont=sf]{subfig}
\usepackage{textcomp}
\usepackage{stfloats}
\usepackage{url}
\usepackage{verbatim}
\usepackage{xcolor}
\usepackage{graphicx}
\usepackage{algorithm}
\usepackage{breakcites}
\usepackage[noend]{algpseudocode}

\usepackage{booktabs}
\usepackage{multirow}
\usepackage{rotating}
\usepackage{adjustbox}
\usepackage[table]{xcolor} 
\usepackage{colortbl}     
\usepackage{longtable,booktabs,array}
\usepackage[colorlinks=true, citecolor=blue]{hyperref}

\def\BibTeX{{\rm B\kern-.05em{\sc i\kern-.025em b}\kern-.08em
    T\kern-.1667em\lower.7ex\hbox{E}\kern-.125emX}}
\usepackage{balance}

\newcounter{mypara} 

\begin{document}
\title{Improving Temporal Consistency and Fidelity \\ at Inference-time  in Perceptual Video Restoration \\by Zero-shot Image-based Diffusion Models}

\author{Nasrin Rahimi,~
        A. Murat~Tekalp,~\IEEEmembership{Life Fellow,~IEEE}

\thanks{Authors are with the Department of Electrical and Electronics Engineering and KUIS AI Center, Koç University, 34450 Istanbul, Turkey. N.R. was partly supported by a KUIS AI Center Fellowship. A.M.T. acknowledges support from  the~Turkish Academy of Sciences (TUBA). Manuscript submitted October~20, 2025.}}

\maketitle

\begin{abstract}
Diffusion models have emerged as powerful priors for single-image restoration, but their application to zero-shot video restoration suffers from  temporal inconsistencies due to the~stochastic nature of sampling and complexity of incorporating explicit temporal modeling. In this work, we address the~challenge of improving temporal coherence in  video restoration using zero-shot image-based diffusion models without retraining or modifying their architecture. We propose two complementary inference-time strategies: 
(1)~Perceptual Straightening Guidance~(PSG)  based on the~neuroscience-inspired perceptual straightening hypothesis, which steers the diffusion denoising process towards smoother temporal evolution by incorporating a curvature penalty in a perceptual space to improve temporal perceptual scores, such as  Fréchet Video Distance~(FVD) and perceptual straightness and (2)~Multi-Path Ensemble Sampling~(MPES), which aims  reducing stochastic variation by ensembling multiple diffusion trajectories to improve fidelity (distortion) scores, such as PSNR and SSIM, without sacrificing sharpness. 
Together, these training-free techniques provide a practical path toward temporally stable high-fidelity perceptual video restoration using large pretrained diffusion models. We performed extensive experiments over multiple datasets and degradation types, systematically evaluating each strategy to understand their strengths and limitations. Our results show while PSG enhances temporal naturalness particularly in case of temporal blur, MPES consistently improves fidelity and spatio-temporal perception–distortion trade-off across all tasks.

\end{abstract}


\section{Introduction}

\IEEEPARstart{T}{he} advent of deep learning has revolutionized the field of image and video restoration. 
Generative models have demonstrated impressive ability to produce perceptually pleasing single-image restoration results. In particular, diffusion models have emerged as state-of-the-art generative priors for image generation and restoration. 
Their strong generative capacity  and stable training makes them attractive for supervised or unsupervised solutions of inverse problems such as image and video super-resolution, deblurring, and inpainting.

Adapting large pretrained image diffusion models to video restoration is particularly appealing, as paired training data for video degradations is often scarce or unavailable. Even when video data is available,  training large high-resolution video generative models is prohibitively expensive. Therefore, a line of zero-shot methods, which apply text-to-image diffusion prior to multiple video restoration tasks without task-specific retraining, have been developed. These task-agnostic, zero-shot large generative methods are valuable for real-world video restoration, where task-specific models may generalize poorly. 
However, applying image-based diffusion models to zero-shot video restoration pose several challenges:

First, zero-shot image diffusion models cannot capture the~spatio-temporal dynamics of degraded videos. Therefore, the stochastic nature of sampling leads to  independent hallucinations in consecutive frames, which results in texture flicker or jitter, and inconsistent motion patterns. A central challenge in video restoration using image-based restoration models is ensuring temporal consistency across frames while maintaining sharpness of each frame. The~apparent trade-off between sharp spatial detail and coherent temporal dynamics has motivated a growing line of research focused on addressing temporal consistency as a core component of video restoration.

Second, most methods that  enforce temporal coherence in zero-shot methods, require architectural modifications or training for added recurrent modules, temporal attention blocks, or motion guidance which are costly. Inference-time strategies, on the other hand, are lightweight and task-agnostic and there is still significant room for improving their ability to directly enforce temporal consistency and naturalness.
Overall, the~key challenge is how to improve temporal coherence while preserving the high spatial quality of pretrained diffusion models, without retraining or sacrificing generality across different degradation types.

To address these challenges, we propose two complementary perspectives: First, the neuroscience-inspired perceptual straightening hypothesis (PSH) \cite{henaff2019perceptual} suggests that natural video sequences which follow curved trajectories in pixel space become straighter when processed by the hierarchical stages of the human visual system. This hypothesis provides a potential proxy for temporal naturalness. Second, the stochastic nature of diffusion sampling causes variations in the outputs across different runs, even with the same conditioning. This motivates variance-reduction strategies such as ensembling, which improves fidelity without retraining.
In this work, we leverage VISION-XL \cite{kwon2024vision} as the baseline, which is a zero-shot video restoration framework based on text-to-image latent diffusion. VISION-XL  applies measurement consistency during sampling; however, the temporal consistency mechanisms used are heuristic leaving room for more principled and broadly applicable inference-time strategies. Our methods can be integrated into VISION-XL or other similar zero-shot pipelines in a coherent manner to enhance temporal stability during inference, without modifying the model architecture.

\bigskip

\noindent\textbf{Contributions.}  
We propose two complementary inference-time optimization strategies that explicitly address temporal inconsistency in zero-shot diffusion video restoration:

\begin{itemize}
    \item \textbf{Perceptual Straightening Guidance (PSG):} Inspired by the neuroscience-based perceptual straightening hypothesis, we introduce a diffusion denoising guidance mechanism that penalizes sequences with large curvature in a perceptual space and encourages frames to evolve more linearly in this embedding space. Evaluation results show that PSG improves perceptual straightness and other temporal metrics, particularly under degradations involving temporal blur and reduces frame-to-frame jitter.
    
    \item \textbf{Multi-Path Ensemble Sampling (MPES):} We propose ensembling multiple diffusion trajectories to reduce the variance of posterior sampling. The proposed MPES offers a tradeoff between sampling from the posterior and MMSE estimation in a controlled manner. We evaluate pixel-space vs. latent fusion strategies. Our results show pixel-space fusion consistently improves both perceptual and distortion metrics across spatial and temporal dimensions -albeit with higher runtime- and significantly improves spatio-temporal perception-distortion trade-off.
\end{itemize}

Together, these contributions demonstrate that lightweight, training-free techniques can substantially improve temporal consistency in zero-shot diffusion video restoration—without retraining or architectural changes.


\section{Related Work}
\subsection{Video Restoration and Temporal Consistency}

Video restoration approaches exploit temporal correlations between frames to improve quality beyond single-image restoration. Early classical methods employed explicit motion estimation and multi-frame modeling \cite{zhao2002super, farsiu2004fast}. Temporal regularization was also  exploited to penalize deviations between consecutive reconstructed frames \cite{takeda2009super} to reduce flicker.  Alternative strategies modeled video as a 3D volume using spatio-temporal priors \cite{protter2008generalizing}. Another approach was using recursive propagation across time \cite{vandewalle2007joint}, though it was prone to error accumulation.

The advent of deep learning has enabled learning rich spatio-temporal representations from training data that lead to significantly enhanced quality. A diverse set of strategies have been employed to achieve temporal coherence. Alignment-based methods such as EDVR \cite{wang2019edvr}, BasicVSR \cite{chan2021basicvsr}, and BasicVSR++ \cite{chan2022basicvsr++} explicitly align features through deformable convolutions and optical flow, while recurrent methods like FRVSR \cite{sajjadi2018frame} encourage temporally consistent results using the previously inferred estimate to restore the subsequent frame. Attention mechanisms including VRT \cite{liang2024vrt}, RVRT \cite{liang2022recurrent} model long-range temporal dependencies through self-attention and space-time attention.

Generative models, such as GANs,  allow optimization of perceptual realism rather than focusing only on pixel-wise fidelity. A GAN-based video super-resolution framework incorporating edge enhancement and motion-compensated multi-frame fusion was proposed in \cite{wang2021video}. TecoGAN \cite{chu2018temporally} pioneered temporally consistent video super-resolution using a spatio-temporal discriminator and ping-pong consistency loss. \cite{rahimi2023spatio} proposes a perceptual VSR model with two discriminators, where a flow discriminator encourages naturalness of motion.

Denoising diffusion probabilistic models (DDPMs)~\cite{ho2020denoising} and variants have recently become state-of-the-art generative approaches for image and video generation. These models have been adapted to specific supervised image and video restoration tasks by conditioning the denoising process on paired (original, degraded) observations. Compared to the GANs, diffusion-based frameworks offer greater stability and typically generate higher-quality results, especially under heavy degradations. Upscale-A-Video~\cite{zhou2024upscale}, a text-guided VSR diffusion model preserves temporal consistency through incorporating temporal layers inside the UNet and VAE-Decoder and global latent propagation mechanism to ensure temporal coherence. DiffVSR \cite{li2025diffvsr} exhibits stable performance against complex degradations through multi-scale temporal alignment. \cite{rota2024enhancing} generate fine-grained details while preserving temporal stability via a trained temporal module. FLAIR \cite{zou2025flair} provides a conditional framework for face video restoration with facial-specific temporal dynamics.
More recent video diffusion methods introduce temporal priors through architectural modifications—such as 3D convolutions or temporal attention~\cite{ho2022video}, or motion-aware modules guided by optical flow~\cite{bao2023latentwarp}.
\vspace{-8pt}

\subsection{Diffusion-Based General Inverse-Problem Solvers (DIS)}
Diffusion-based inverse problem solvers (DIS) exploit pre-trained diffusion models to capture the the prior image distribution \( p_\theta(x) \). Given a degradation model, DIS methods approximate the posterior \( p_\theta(x|y) \), where \( y \) is a degraded observation. This is achieved by combining the data prior \( p_\theta(x) \) with the likelihood \( p(y|x) \), enabling iterative sampling guided by the measurement \( y \). In the video domain, diffusion models face additional challenges due to the temporal dimension. 

Naïvely applying image-trained diffusion models frame-by-frame often leads to flicker and motion inconsistency. 
To address this, DIS-based video restoration approaches~\cite{daras2024warped, kwon2024solving} employ explicit strategies to control temporal consistencies. 
DIS-based approaches have been successfully applied to a wide range of tasks, including deblurring, super-resolution, inpainting, colorization, and compressed sensing.
ZVRD \cite{cao2025zero} uses a pre-trained single image diffusion model for video in a zero-shot setting by mechanisms like short/long-range temporal attention, consistency guidance, and spatio-temporal noise coordination to achieve temporally coherent restoration.
DiffIR2VR-Zero \cite{yeh2024diffir2vr} performs zero-shot temporally consistent video restoration by using any pre-trained image diffusion model. It combines hierarchical latent warping  with hybrid token merging and maintains temporal consistency while preserving restoration quality for different restoration tasks.
VISION-XL \cite{kwon2024vision} a Zero-shot video restoration model utilizes latent diffusion model with pseudo-batch consistent sampling and data consistency refinement to improve temporal coherence and fidelity.

Despite recent progress, preserving temporal coherence in zero-shot video restoration remains challenging. Many existing methods rely on retraining or architectural changes, while general inference-time strategies for temporal consistency are still limited. We address this gap by proposing inference-time (training-free) methods to improve temporal consistency and fidelity without modifying the underlying diffusion model.
\vspace{-8pt}

\subsection{Perceptual Straightening Hypothesis (PSH)}

The perceptual straightening hypothesis (PSH), introduced by neuroscientist Hénaff, et al.~\cite{henaff2019perceptual}, states that the human visual system transforms natural video sequences to a perceptual feature space, where motion follows straighter trajectories compared to the pixel-intensity domain. 
 In contrast, unnatural or inconsistent visual sequences tend to exhibit greater curvature in this perceptual feature space.

The PSH has inspired several works in video processing. Kancharla and Channappayya~\cite{kancharla2021frameprediction} used PSH in a video frame prediction model and showed improvement in visual quality by minimizing curvature of the trajectory in the perceptual space. They also proposed a no-reference quality measure  based on PSH  to evaluate naturalness of motion in user-generated videos~\cite{kancharla2021ugcqa}.
In our earlier work~\cite{rahimi2023spatio}, we employed perceptual straightness as an evaluation metric for temporal naturalness in video super-resolution (VSR).
More recently, Internò et al.~\cite{interno2025ai} applied the PSH to detect AI-generated videos, showing that unnatural temporal dynamics introduced by generative models result in higher perceptual curvature, which can serve as a discriminative signal.

In this work, we employ perceptual straightness as an optimization criterion at inference-time to enforce perceptual straightness of videos generated by zero-shot VISION-XL pipeline. We show the proposed method can avoid unnatural or inconsistent restored sequences, straightening their perceptual trajectory without explicit motion supervision.

\subsection{Ensembling in Generative Models}
\label{subsec: rw_ensemble}
 
Due to the stochastic nature of diffusion models,
each time the diffusion model processes the same input, the noisy latent follows a slightly different trajectory because of the randomness in the denoising process. 
This trajectory variability can cause small differences in fine-grained image details, as well as inconsistencies in temporal dynamics of the video. 
The iterative nature of diffusion sampling can make the situation worse, where small errors accumulate at each denoising step.

Ensembling methods combine multiple models or 
predictions to reduce variance, and improve robustness for a variety of machine learning and generative tasks. In EnsembleGAN~\cite{zhang2019ensemblegan} and Dropout-GAN~\cite{mordido2018dropout},  ensembling techniques, such as combining multiple discriminators or generators, have been explored to stabilize training and improve visual fidelity. In diffusion models, prior works have explored multiple samplings for  uncertainty estimation~\cite{dhariwal2021diffusion}, classifier-free guidance tuning~\cite{ho2022classifier}, and best-of-$N$ selection~\cite{saharia2022photorealistic}. eDiff-I \cite{zhang2025artist} constructs ensembles specialized for different synthesis stages to enhance text alignment in text-to-image generation. Adaptive Feature Aggregation (AFA) \cite{wang2025ensembling} proposes dynamically merging multiple diffusion models' feature outputs using a spatial-aware attention mechanism, improving generation robustness. Recently,  Korkmaz et al. \cite{korkmaz2025leveraging} applied ensembling to diffusion-based image super-resolution by generating multiple  outputs, ranking them using a vision-language model (VLM), and fusing the top-ranked ones via pixel-wise averaging.

These studies motivate us to investigate multi-path ensemble sampling in zero-shot image-based video restoration, where improving fidelity and temporal consistency are critical.

\section{Baseline: VISION-XL Video Inverse Problem Solver}
\label{sec:vision_xl_zero_shot}

\textit{VISION-XL}~\cite{kwon2024vision} is a state-of-the-art video inverse problem solver that adapts pretrained SDXL~\cite{podell2023sdxl} for zero-shot video restoration. It introduces algorithmic strategies to extend single-frame latent diffusion models to multi-frame video tasks.

Given a degraded video $\{\mathbf{y}^{[n]}\}_{n=1}^N$ and degradation operator $A: \mathbb{R}^{H \times W \times 3} \to \mathbb{R}^{H' \times W' \times 3}$, the goal is to restore clean frames $\{\mathbf{x}_0^{[n]}\}_{n=1}^N$.

One conditioning strategy used in VISION-XL is initializing the diffusion sampling process using DDIM inversion~\cite{song2020denoising} of the measurements. Instead of starting from pure noise, VISION-XL applies DDIM inversion to the latent representation of the first measurement frame, and uses the resulting noisy latent $\mathbf{z}_{\tau}$ to initialize all frame latents by repetition $\mathbf{Z}_{\tau} := [\mathbf{z}_{\tau}, \ldots, \mathbf{z}_{\tau}]$.

Starting the denoising process from timestep $\tau$, at each timestep $t \in [\tau, \ldots, 0]$, the noisy latent $\mathbf{z}_{t}$ is passed to SDXL UNet to predict noise $\boldsymbol{\epsilon}_\theta(\mathbf{z}_t, t)$, then the clean latent $\hat{\mathbf{z}}_0^{(t)}$ is predicted using Tweedie’s formula  in \ref{eq: tweedie's formula}.

In Data Consistency Optimization step, the latent $\hat{\mathbf{z}}_0^{(t)}$  is decoded using VAE decoder and are
refined by minimizing a data-consistency loss using $N_{GC}$-step conjugate gradient (CG) optimization:
\begin{equation}
\bar{\mathbf{x}}_0 := \arg\min_{X \in \hat{X}_0 + \mathcal{K}_l} \left\| Y - A(X) \right\|_2^2
\label{eq:visionxl_data_consistency_projection}
\end{equation}

To mitigate VAE errors and suppress high-frequency artifacts, a Gaussian low pass filter is applied on refined frames and then they are encoded to latent space and are re-noised into $\mathbf{z}_{t-1}$ to continue the sampling.
\begin{equation}
\begin{split}
\bar{\mathbf{z}}_0 &= E_\phi\left( \bar{\mathbf{x}}_0 \right)  \\
\mathbf{z}_{t-1} &= \sqrt{\bar{\alpha}_{t-1}} \bar{\mathbf{z}}_0+ \sqrt{1 - \bar{\alpha}_{t-1}} \cdot \boldsymbol{\epsilon}_{t},
\end{split}
\end{equation}
where $\boldsymbol{\epsilon}_{t}$ is composed of batch-consistent noise and deterministic noise:
\begin{equation}
\begin{split}
c_1 &= \sqrt{1 - \bar{\alpha}_{t-1}} \cdot \eta \\
c_2 &= \sqrt{1 -\bar{\alpha}_{t-1}} \cdot \sqrt{1 - \eta^2} \\
\mathbf{z}_{t-1} &= \sqrt{\bar{\alpha}_{t-1}} \bar{\mathbf{z}}_0+ c_1 \cdot \boldsymbol{\epsilon}_{\text{fixed}} + c_2 \cdot \boldsymbol{\epsilon}_\theta(\mathbf{z}_t, t)
\label{eq:visionxl_renoising}
\end{split}
\end{equation}

\noindent\textbf{Temporal Consistency Mechanisms.}
VISION-XL uses several inference-time strategies to enforce temporal consistency:

\begin{itemize}
    \item Consistent DDIM Initialization: First-frame DDIM inversion is reused across frames.
    \item Synchronized Sampling: Shared noise schedule and conditioning across frames.
    \item Low-Pass Filtering: Reduces flickering by suppressing high-frequency noise.
    \item CG Optimization: Joint data consistency improves coherence under temporal degradation.
\end{itemize}

\noindent\textbf{Limitations.}  
Although VISION-XL's inference-time strategies improve temporal coherence and reduce flickering, there exists some limitations: shared latent initialization can over-smooth motion, there is no inter-frame communication during sampling, and CG updates lack explicit temporal modeling. While stronger temporal priors have been proposed in other diffusion frameworks, they often require architectural modifications or retraining on large video datasets—making them incompatible with zero-shot settings.

Temporal inconsistencies in zero-shot diffusion models are also caused by perceptual instability and stochastic variation during sampling. This motivates our work towards stronger temporal coherence: In the following, we propose a set of lightweight, inference-time strategies designed to improve the temporal coherence of zero-shot diffusion-based video restoration models, such as VISION-XL, without sacrificing sample quality or generality.

\section{Perceptual Straightening Guidance (PSG)}

\subsection{Motivation and Objective}

A major limitation of applying image-based diffusion models to video restoration is the potential of temporal incoherence. While the VISION-XL framework achieves state-of-the-art performance in zero-shot video inverse problems, it primarily focuses on data consistency and lacks an explicit mechanism to enforce temporal consistency. This is important, as the human visual system exhibits strong sensitivity to temporal artifacts.
We propose using the PSH as a guiding principle to reduce unnatural fluctuations in the video while preserving sharpness. To this effect, we introduce Perceptual Straightening Guidance (PSG), an inference-time optimization strategy, where video samples obtained by the reverse diffusion process are further optimized using perceptual straightness as a criterion.
Unlike changes to model architecture or retraining, PSG is lightweight and task-agnostic, making it suitable across different video degradations in a zero-shot setting.

\subsection{Formulation of PSG}
In order to guide the diffusion  process toward restoring frames that follow straighter trajectories in the perceptual representation space, we incorporate perceptual straightening as a refinement step during VISION-XL sampling. This involves defining a perceptual straightening loss that penalizes sequences with large curvature in perceptual space, encouraging the generation of temporally natural frames aligned with motion patterns as perceived by the human visual system.

\noindent\textbf{Perceptual Representation Computation}
Given a sequence of $N$ refined frames $\{\bar{x}_0^{[n]}\}_{n=1}^N$ obtained from the VISION-XL sampling process at timestep $t$, we compute hierarchical perceptual representations at different stages of the visual processing pipeline. To extract the perceptual representation for a video sequence, we use the two-stage nonlinear model (RetinalDN and V1) suggested by~\cite{henaff2019perceptual} that simulates the early processing of human visual system.
\begin{align}
R_{\text{pixel}} &= \{\bar{x}_0^{[n]}\}_{n=1}^N \\
R_{\text{retina}} &= \{f_{\text{retina}}(\bar{x}_0^{[n]}\}_{n=1}^N \\
R_{\text{V1}} &= \{f_{\text{V1}}(f_{\text{retina}}(\bar{x}_0^{[n]}\}_{n=1}^N
\end{align}
where $f_{\text{retina}}(\cdot)$ represents the RetinalDN transformation and $f_{\text{V1}}(\cdot)$ denotes the steerable pyramid processing that simulates V1 complex cell responses. The details  can be found in Appendix \ref{apx: PSH}. As illustrated in Figure~\ref{fig:PS}, each representation forms a trajectory in high-dimensional space, where temporal consistency is measured through trajectory curvature.

\begin{figure}[t]
	\centering
	\includegraphics[width=0.8\columnwidth]{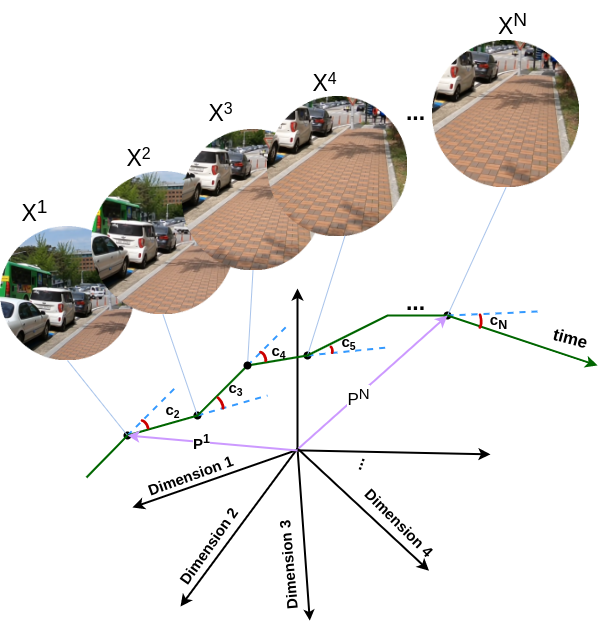}
	\caption{\small  Visualization of a high-dimensional perceptual representation of a video sequence and the concept of curvature}
	\label{fig:PS}
\end{figure}

\noindent\textbf{Curvature Computation }

Given perceptual embeddings $P^n \in R_{\text{V1}}$ for frame $n$, we define displacement vectors as $V^n = P^n - P^{n-1}$. The curvature at frame $n$ is computed as the angle between successive displacement vectors between frame representations in V1 space:
\begin{equation}
\text{Curv}(n) = \arccos\left(\frac{V^n \cdot V^{n+1}}{\|V^n\| \|V^{n+1}\|}\right)
\end{equation}
where the dot product measures the similarity between consecutive displacement directions. The overall trajectory curvature is obtained by averaging across the sequence:
\begin{equation}
\text{Curv}(R_{V1}) = \frac{1}{N-2} \sum_{n=2}^{N-1} \text{Curv}(n)
\end{equation}

This high-dimensional curvature computation detect subtle temporal inconsistencies that may not be visible in pixel space and provides a more perceptually meaningful measure of temporal naturalness \cite{rahimi2023spatio}.

\noindent\textbf{Perceptual Straightening Penalty Formulation}  
The goal of PSG is to penalize video sequences with large V1 curvature, thereby encouraging smoother temporal trajectories in the perceptual space. We formulate the perceptual straightening penalty $\mathcal{L}_{PS}$ as a differentiable guidance term that can be seamlessly integrated into the VISION-XL inference pipeline. Given a sequence of predicted clean frames $\bar{X_0} = \{\bar{x}_0^{[n]}\}_{n=1}^N$ at timestep $t$ during the reverse diffusion process, the penalty is defined as:
\begin{equation}
\mathcal{L}_{\text{PS}}(\bar{X_0}) = \text{ReLU}(\text{Curv}(f_{\text{V1}}(f_{\text{retina}}(\bar{X}_0))))
\end{equation}
the ReLU activation is used primarily for numerical stability. Since curvature values are inherently non-negative (angles between consecutive displacement vectors range from 0° to 180°), its practical impact is minimal.

\begin{figure*}[t]
	\centering
	\includegraphics[width=0.8\textwidth]{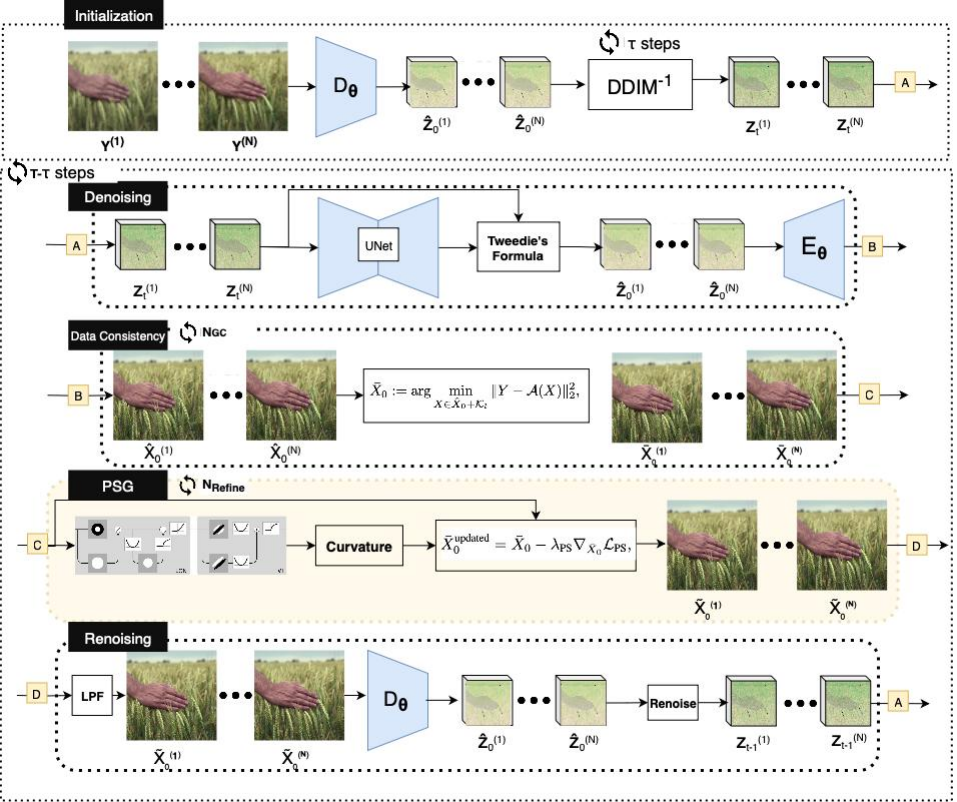}
	\caption{\small  The proposed PSG block within the inference process of VISION-XL.}
	\label{fig: PSG}
\end{figure*}

We incorporate the perceptual straightening penalty as guidance during diffusion sampling. As illustrated in Figure \ref{fig: PSG}, at each denoising timestep after predicting the clean sequence, we iteratively compute \( \mathcal{L}_{\text{PS}} \) and backpropagate the penalty computed in the perceptual space to refine the predicted clean frames using gradient descent:
\begin{equation}
\bar{X}_0^{\text{updated}} = \bar{X}_0 - \lambda_{\text{PS}} \nabla_{\bar{X}_0} \mathcal{L}_{\text{PS}},
\end{equation}
where \( \lambda_{\text{PS}} \) is a tunable guidance weight controlling the influence of the perceptual straightening penalty during sampling. 

The details of PSG process is described in Algorithm~\ref{alg:perceptual_refinement}.

\begin{algorithm}[H]
\caption{Perceptual Straightness Refinement (PSG)}
\label{alg:perceptual_refinement}
\begin{algorithmic}[1]
\Require Input frames $\mathbf{\bar{X}}_{0}$, number of iterations $N_{\text{refine}}$
\Ensure Perceptually refined frames $\tilde{\mathbf{X}}_0$
\State $\tilde{\mathbf{X}}_0 \leftarrow \mathbf{\bar{X}}_{0}.\text{detach}().\text{clone}().\text{requires\_grad\_}(\text{True})$
\For{$i = 1$ \textbf{ to } $N_{\text{refine}}$}
    \State $\mathcal{L}_{PS} \leftarrow \text{PerceptualStraightnessLoss}(\tilde{\mathbf{X}}_0)$
    \State $\nabla \mathcal{L}_{PS} \leftarrow \text{autograd.grad}(\mathcal{L}_{PS}, \tilde{\mathbf{X}}_0)$
    \State $\tilde{\mathbf{X}}_0 \leftarrow \tilde{\mathbf{X}}_0 - \lambda_{\text{PS}} \cdot \nabla \mathcal{L}_{PS}$
    \If{$(i + 1) \bmod 2 = 0$}
        \State $\lambda_{\text{PS}} \leftarrow \lambda_{\text{PS}} \times 0.8$ 
    \EndIf
\EndFor
\State \Return $\tilde{\mathbf{X}}_0$
\end{algorithmic}
\end{algorithm}

This regularization steers the model toward generating temporally smoother aligned outputs without requiring motion supervision or reference video data.


\section{Multi-Path Ensemble Sampling (MPES)}
\subsection{Motivation and Objective}

The inherent stochasticity of diffusion sampling, where each denoising run can yield slightly different results even for the same input, introduces variations across different  trajectories.
Although individual predictions may be noisy, their average is likely to be more accurate statistically, and leads to better approximation of the ground truth~\cite{efron2011tweedie, bishop2006pattern}. 
In the proposed multi-path ensemble sampling. instead of relying on a single diffusion trajectory, we generate multiple sampling paths and fuse the resulting video samples to achieve better results.

\subsection{Sources of Stochasticity}
In Vision-XL (baseline), repeated runs with the same input can yield different results due to multiple sources of randomness. First,  injected noise causes denoising paths to diverge, leading the U-Net to produce different predictions over time—even for identical inputs—resulting in multiple valid sampling trajectories.
Second, since Vision-XL operates in the latent space via a pretrained VAE, small input differences introduce varying reconstruction errors. These errors can accumulate during iterative encode-decode cycles and drift the latent from the clean manifold.
Lastly, the data consistency step involves conjugate gradient optimization, which can follow different paths depending on the input. 

These trajectories explore different regions of the data distribution through diffusion prior.  This diversity can be leveraged through ensembling to yield more robust and temporally consistent reconstructions.

\subsection{Proposed Framework}
\label{subsec: ensemble_stratefies}
We propose a multi-path ensemble sampling framework that generates multiple diffusion trajectories and fuses them to obtain robust video reconstruction. The key idea is to exploit the statistical advantages of ensemble methods while preserving the temporal consistency of video restoration results.

Let $\mathcal{S}_\theta$ denote a diffusion-based inverse problem solver (such as VISION-XL) that takes degraded measurements $\mathbf{Y}$  as input and generates a restored video sequence $\mathbf{X}$. Our multi-path ensemble approach, summarized in Algorithm \ref{alg:independent_sampling}, produces $K$ independent sampling trajectories:
\begin{equation}
\mathbf{X}^{(k)} = \mathcal{S}_\theta(\mathbf{Y}, \boldsymbol{\xi}^{(k)}), \quad k = 1, 2, \ldots, K
\label{eq:multi_path_sampling}
\end{equation}
where $\boldsymbol{\xi}^{(k)}$ denotes the random seed or stochastic components unique to the $k$-th path.

\begin{figure}[t]
	\centering
	\includegraphics[width=0.95\columnwidth]{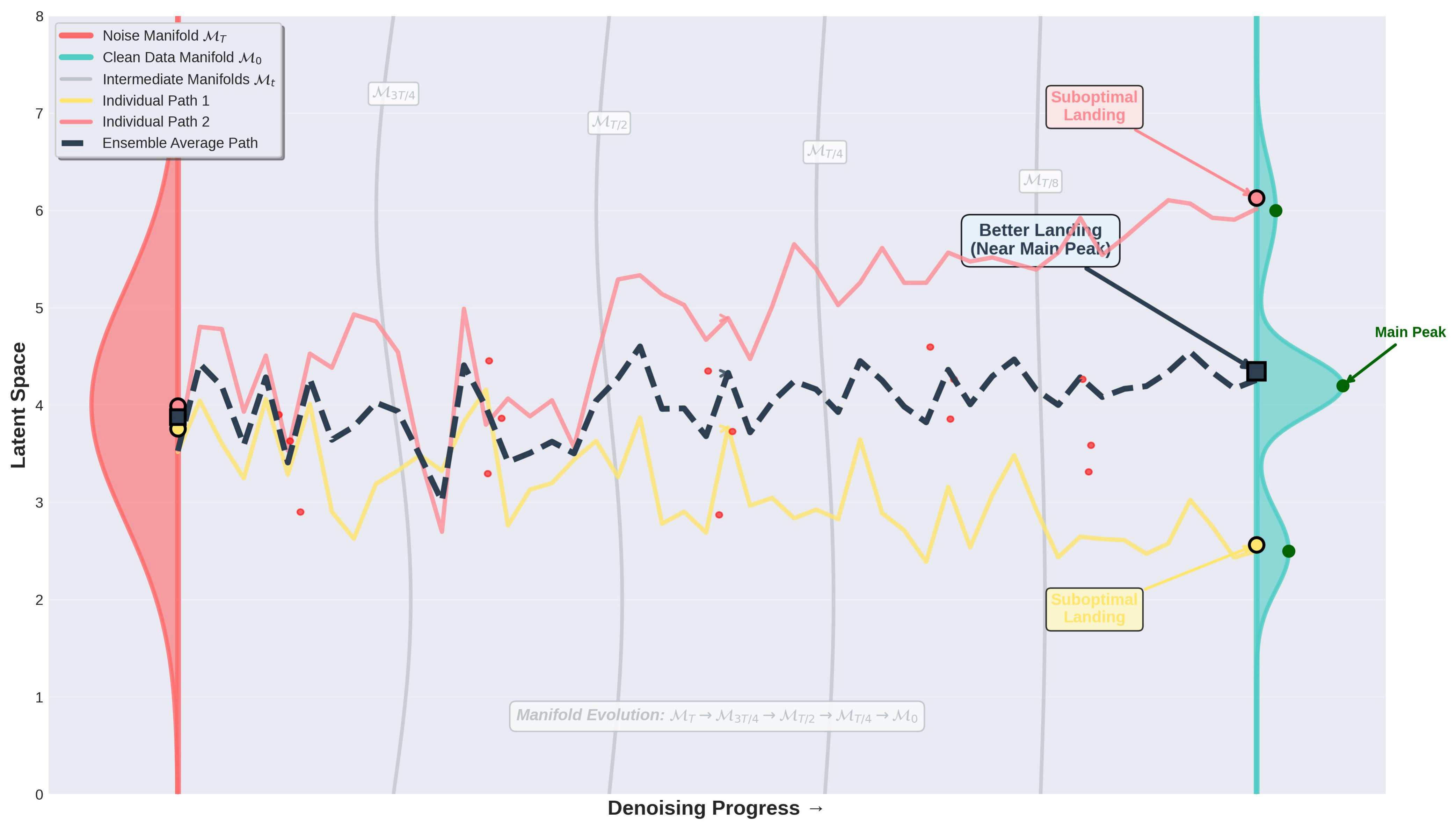} \vspace{-5pt}	

\caption{\small Multi-Path Ensemble Sampling: Individual vs. average trajectories.  The average leads to a better solution than individual ones.}
	\label{fig:ind_ensemble}
\end{figure}

In this method, each sampling trajectory $\{\mathbf{z}_t^{(k)}\}_{t=0}^\tau$ creates its own independent path through the latent manifold. As a result, the paths can diverge significantly during the sampling process and may explore different regions of the posterior distribution $p(\mathbf{X}|\mathbf{Y})$ (Figure \ref{fig:ind_ensemble}).

The final ensemble prediction is obtained by using  a fusion function $\mathcal{F}$ :
\begin{equation}
\mathbf{X}_{ensemble} = \mathcal{F}(\mathbf{X}^{(1)}, \mathbf{X}^{(2)}, \ldots, \mathbf{X}^{(K)})
\label{eq:ensemble_fusion}
\end{equation}

\begin{algorithm}[t]
\caption{Multi-Path Ensemble Sampling}
\label{alg:independent_sampling}
\begin{algorithmic}[1]
\Require{Measurements $\mathbf{Y}$, Number of paths $K$}
\For{$k = 1, \ldots, K$}
   \State $\mathbf{z}_\tau^{(k)} \leftarrow \text{BatchConsistentInversion}(\mathbf{Y}, \boldsymbol{\xi}^{(k)})$
   \For{$t = \tau, \tau-1, \ldots, 1$}
       \State $\hat{\mathbf{z}}_0^{(k)} \leftarrow \text{TweedieDenoising}(\mathbf{z}_t^{(k)}, t)$ 
       \State $\mathbf{\hat{X}}_0^{(k)} \leftarrow \text{VAEDecode}(\hat{\mathbf{z}}_0^{(k)})$
       \State $\mathbf{\bar{X}}_0^{(k)} \leftarrow \text{DataConsistency}(\mathbf{\hat{X}}_0^{(k)}, \mathbf{Y})$ 
       \State $\mathbf{z}_{t-1}^{(k)} \leftarrow \text{Renoise\&Encode}(\mathbf{\bar{X}}_0^{(k)}, \boldsymbol{\epsilon}^{(k)}_t)$ 
   \EndFor
\EndFor
\State \Return $\mathcal{F}_{final}(\mathbf{\bar{X}}_0^{(1)}, \ldots, \mathbf{\bar{X}}_0^{(K)})$
\end{algorithmic}
\end{algorithm}

\noindent\textbf{Fusion Strategies}
To combine outputs from multiple diffusion sampling trajectories, we consider different representation space in which fusion is done.   

\noindent\paragraph{\textbf{Representation Spaces}}
Fusion can be applied at different stages of the generation process:

\begin{itemize}
    \item Latent-Space Fusion: Predicted latents from each trajectory are fused in latent space.
    \item Pixel-Space Fusion: Latents are decoded into images first, and fusion is done in the pixel space.
\end{itemize}
In all the cases uniform averaging is used where all trajectories contribute equally to the final output.

\subsection{Properties of Multi-Path Ensemble Sampling}
\label{subsec: ensemble_theory}
Ensembling multiple diffusion trajectories offers the following theoretical advantages: 
\vspace{6pt}

\noindent\textbf{Manifold Regularization via Trajectory Averaging}
As illustrated in Figure \ref{fig:ind_ensemble}, diffusion sampling can be considered  as a trajectory across a sequence of latent manifolds $\mathcal{M}_t$, starting from pure noise $\mathcal{M}_T$ and moving toward the clean data manifold $\mathcal{M}_0$. Each sample path $\{\mathbf{z}_t^{(k)}\}_{t=0}^T$ may deviate from the related manifold because of the accumulated noise or optimization errors.

Averaging latent from multiple trajectories at each timestep (manifold) has a regularizing effect: when individual latents are near the manifold, their Euclidean average will also stay near the manifold.
\vspace{6pt}

\noindent\textbf{Posterior Smoothing and MMSE Estimation}
As discussed in Appendix~\ref{subsec:diffusion_inverse_problem}, diffusion-based inverse solvers approximate sampling from the posterior distribution $p(\mathbf{X}|\mathbf{Y})$.  Under squared loss, the optimal Bayesian estimator called the Minimum Mean Squared Error (MMSE) estimator is the posterior mean:
\begin{equation}
\mathbf{X}_{MMSE} = \mathbb{E}[\mathbf{X}|\mathbf{Y}] = \int \mathbf{X} \, p(\mathbf{X}|\mathbf{Y}) \, d\mathbf{X}
\label{eq:mmse_estimator}
\end{equation}

This estimator achieves a reconstruction with the minimum expected squared difference from the ground truth. For high-dimensional distributions that the exact estimation of this expectation is not possible, it can be approximated using Monte Carlo sampling \cite{robert2004monte}:
\begin{equation}
\mathbf{X}_{MMSE} \approx \frac{1}{K} \sum_{k=1}^K \mathbf{X}^{(k)}, \quad \mathbf{X}^{(k)} \sim p(\mathbf{X}|\mathbf{Y})
\label{eq:mc_approximation}
\end{equation}

According to the central limit theorem, the approximation error which is the standard deviation of $\mathbf{X}^{(k)}$ decreases as $O(1/\sqrt{K})$.
This analysis offers a foundation for understanding why multi-path ensemble sampling  improves quality of video restoration.

\section{Evaluation: Experimental Results}

\subsection{Evaluation Setup}
\label{subsec:evaluation-setup}
\noindent\textbf{Datasets}
We evaluate our methods on two benchmark datasets: DAVIS and REDS4. We use 73 sequences from DAVIS-2017~\cite{Pont-Tuset_arXiv_2017}, resized to 768$\times$1280, preserving their original landscape orientation. DAVIS includes high-quality videos with challenging scenarios, suitable for spatio-temporal restoration. REDS4 consists of four landscape-oriented clips from the~REDS dataset~\cite{NTIRE2019}, each with 100 frames at 720$\times$1280 resolution. It offers diverse motion patterns and scene complexities, which is complementary to DAVIS.
\vspace{6pt}

\noindent\textbf{Degradation Models}
We evaluate on five spatio-temporal inverse problems formulated as compositions of spatial and temporal operators. Spatial degradations include: Super-Resolution (4× downsampling) and Deblurring (Gaussian blur with $\sigma=3.0$). Spatio-temporal degradations include:  SR+ and Deblur+ (each adding 7-frame temporal averaging on top of their spatial counterpart) and Temporal Blur (13-frame averaging). The measurement model is defined as $\mathbf{y} = \mathcal{A}(\mathbf{x})$.
\vspace{6pt}

\noindent\textbf{Evaluation Metrics}
To assess fidelity and perceptual quality, we use spatial distortion metrics (PSNR\cite{hore2010image},  SSIM \cite{wang2004image}), spatial perceptual metrics (LPIPS\cite{zhang2018unreasonable}, NIQE\cite{mittal2012making}), temporal metrics (OF-MSE, FVD \cite{unterthiner2019fvd}, Perceptual Straightness) and spatio-temporal metrics (P-ST, and D-ST\cite{rahimi2023spatio}). 
\vspace{6pt}

\noindent \textbf{Implementation Platform}
All experiments are conducted on a single NVIDIA V100 GPU with 32GB memory. Also, all inferences are done with mixed precision computation using half precision (torch.float16) for memory efficiency.


\subsection{Effect of Perceptual Straightening Guidance (PSG)}
\label{subsec:ablation-perceptual-straightening}

We implemented a “Sequential Optimization" approach comprised of a two-stage optimization pipeline, where data consistency refinement precedes perceptual straightening optimization.
In each sampling step of the VISION-XL pipeline, after decoding the estimated clean latent sequence to pixel-space, we first apply $N_{\text{DDS}} = 5$ conjugate gradient iterations for data consistency (DDS), then we compute perceptual straightening loss and apply $N_{\text{PS}} = 10$ gradient descent steps to enforce straightness in the perceptual space with the~learning rate $\lambda_{\text{PS}} = 0.5$.

We also explored alternative PSG schemes, such as  “Alternating Optimization," where we alternate single steps of data consistency and perceptual straightening, and  “Joint Optimization," where we simultaneously optimize both objectives.
However,  neither scheme lead to better results. We observed that minimizing curvature  when frames are still far from the~natural-image manifold acts like a temporal low-pass filter to reduce frame-to-frame variation (i.e., introduces temporal blurring) instead of preserving motion-consistent changes.  Overall, applying PSG optimization before the frames are sufficiently “clean” and aligned may not produce  perceptually natural motion. 
\vspace{6pt}

\noindent\textbf{Results and Analysis}

We report quantitative results for the proposed sequential optimization in Table \ref{tab:PS_comparison_DAVIS} and \ref{tab:PS_comparison_REDS4}. 
Across DAVIS and REDS4 and for all degradations, the \textit{Straightness} score improves compared to \textit{Baseline+}, evidence that using Perceptual straightening guidance straightens the V1-trajectory for the restored videos. We observe that the spatial scores, PSNR, SSIM, and LPIPS do not improve but stay on par with Baseline+ indicating that perceptual straightening optimization is primarily a temporal regularizer and affects temporal scores.

Our analysis reveals a dependency between task complexity and effectiveness of perceptual straightening. When degradations contain temporal blur (\textbf{sr+}, \textbf{deblur+}, \textbf{temporal-blur}), improvements in the Straightness metric is more pronounced and Straightness correlates strongly with the FVD across both datasets. One explanation for this  is that Straightness specifically measures the smoothness of frame-to-frame evolution in a perceptual space, and our perceptual straightening guidance primarily improves temporal coherence. In the motion-blur degradation, the dominant error is temporal (suppressed or smeared motion cues); correcting it both straightens the perceptual trajectory and reduces FVD. 
In contrast, for spatial-only degradations, temporal dynamics are largely preserved; gains in Straightness reflect minor flicker reduction but do not address the spatial realism that mainly determines FVD; then, the two metrics are not strongly correlated.

\begin{table*}[htbp]
\centering
\caption{Quantitative evaluation on \textbf{DAVIS} dataset comparing \textbf{Perceptual Straightening} optimization versus \textbf{Baseline+} across various degradation tasks. Best values are \textbf{bolded}.}
\label{tab:PS_comparison_DAVIS}
\resizebox{\textwidth}{!}{%
\begin{tabular}{|c|c|cccc|cccc|cc|}
\hline
& & \multicolumn{4}{c|}{\textbf{Spatial}} & \multicolumn{4}{c|}{\textbf{Temporal}} & \multicolumn{2}{c|}{\textbf{Spatio-Temporal}} \\
\cline{3-12}
\textbf{Degradation} & \textbf{Method} & \textbf{PSNR↑} & \textbf{SSIM↑} & \textbf{NIQE↓} & \textbf{LPIPS↓} & \textbf{flow\_MSE↓} &  \textbf{fvd-videogpt↓} & \textbf{fvd-styleganv↓} & \textbf{Straightness↑} & \textbf{D-ST↓} & \textbf{P-ST↓} \\
\hline
\cellcolor[HTML]{D3D3D3} & \cellcolor[HTML]{F0F0F0}PS & 29.6799 & 0.8271 & \textbf{6.0284} & \textbf{0.2223} & \textbf{0.0142} & 85.9838 & 83.3062 & \textbf{95.760} & 159.9752 & \textbf{0.1330} \\
\multirow{-2}{*}{\cellcolor[HTML]{D3D3D3}\textbf{sr}} & \cellcolor[HTML]{F0F0F0}Baseline+ & \textbf{29.7367} & \textbf{0.8294} & 6.1289 & 0.2249 & 0.0145 & \textbf{83.8501} & \textbf{82.2478} & 95.454 & \textbf{159.4358} & 0.1350 \\
\hline
\cellcolor[HTML]{D3D3D3} & \cellcolor[HTML]{F0F0F0}PS & 28.5560 & 0.7871 & \textbf{6.2095} & \textbf{0.2861} & 0.0151 & \textbf{117.1282} & \textbf{116.1784} & \textbf{92.790} & 178.4411 & \textbf{0.1766} \\
\multirow{-2}{*}{\cellcolor[HTML]{D3D3D3}\textbf{+sr}} & \cellcolor[HTML]{F0F0F0}Baseline+ & \textbf{28.5687} & \textbf{0.7878} & 6.2499 & 0.2869 & \textbf{0.0151} & 122.4390 & 120.0357 & 92.322 & \textbf{178.1407} & 0.1781 \\
\hline
\cellcolor[HTML]{D3D3D3} & \cellcolor[HTML]{F0F0F0}PS & 31.1147 & 0.8525 & 6.6008 & 0.2303 & 0.0126 & 70.8423 & 70.7495 & \textbf{95.940} & 130.8105 & \textbf{0.1376} \\
\multirow{-2}{*}{\cellcolor[HTML]{D3D3D3}\textbf{deblur}} & \cellcolor[HTML]{F0F0F0}Baseline+ & \textbf{31.1342} & \textbf{0.8529} & \textbf{6.5927} & \textbf{0.2302} & \textbf{0.0123} & \textbf{69.1178} & \textbf{68.6272} & 95.616 & \textbf{130.3061} & 0.1379 \\
\hline
\cellcolor[HTML]{D3D3D3} & \cellcolor[HTML]{F0F0F0}PS & \textbf{28.1387} & \textbf{0.7736} & 6.6618 & \textbf{0.3471} & 0.0159 & \textbf{164.5605} & \textbf{164.5999} & \textbf{93.096} & 189.0676 & \textbf{0.2136} \\
\multirow{-2}{*}{\cellcolor[HTML]{D3D3D3}\textbf{+deblur}} & \cellcolor[HTML]{F0F0F0}Baseline+ & 28.1199 & 0.7735 & \textbf{6.6476} & 0.3473 & \textbf{0.0159} & 169.0877 & 167.1298 & 92.25 & \textbf{189.5531} & 0.2157 \\
\hline
\cellcolor[HTML]{D3D3D3} & \cellcolor[HTML]{F0F0F0}PS & 30.4432 & 0.7938 & \textbf{4.5684} & 0.2151 & \textbf{0.0127} & \textbf{49.5639} & \textbf{48.8714} & \textbf{94.410} & \textbf{159.1802} & \textbf{0.1305} \\
\multirow{-2}{*}{\cellcolor[HTML]{D3D3D3}\textbf{temporal blur}} & \cellcolor[HTML]{F0F0F0}Baseline+ & \textbf{30.4446} & \textbf{0.7938} & 4.5720 & \textbf{0.2149} & 0.0127 & 51.4089 & 50.0101 & 94.194 & 159.1931 & 0.1308 \\
\hline
\end{tabular}%
}
\end{table*}

\begin{table*}[htbp]
\centering
\caption{Quantitative evaluation on \textbf{REDS4} dataset comparing \textbf{Perceptual Straightening} optimization versus \textbf{Baseline+} across various degradation tasks. Best values are \textbf{bolded}.}
\label{tab:PS_comparison_REDS4}
\resizebox{\textwidth}{!}{%
\begin{tabular}{|c|c|cccc|cccc|cc|}
\hline
& & \multicolumn{4}{c|}{\textbf{Spatial}} & \multicolumn{4}{c|}{\textbf{Temporal}} & \multicolumn{2}{c|}{\textbf{Spatio-Temporal}} \\
\cline{3-12}
\textbf{Degradation} & \textbf{Method} & \textbf{PSNR↑} & \textbf{SSIM↑} & \textbf{NIQE↓} & \textbf{LPIPS↓} & \textbf{flow\_MSE↓} &  \textbf{fvd-videogpt↓} & \textbf{fvd-styleganv↓} & \textbf{Straightness↑} & \textbf{D-ST↓} & \textbf{P-ST↓} \\
\hline
\cellcolor[HTML]{D3D3D3} & \cellcolor[HTML]{F0F0F0}PS & 26.2507 & 0.7413 & \textbf{5.2627} & \textbf{0.3130} & \textbf{0.0131} & 54.0367 & 52.8803 & \textbf{108.644} & 178.3603 & \textbf{0.1730} \\
\multirow{-2}{*}{\cellcolor[HTML]{D3D3D3}\textbf{sr}} & \cellcolor[HTML]{F0F0F0}Baseline+ & \textbf{26.2763} & \textbf{0.7431} & 5.3510 & 0.3195 & 0.0135 & \textbf{51.6672} & \textbf{51.8888} & 103.626 & \textbf{177.7404} & 0.1767 \\
\hline
\cellcolor[HTML]{D3D3D3} & \cellcolor[HTML]{F0F0F0}PS & 25.1730 & 0.6873 & \textbf{5.8042} & \textbf{0.3737} & \textbf{0.0147} & \textbf{140.0593} & \textbf{139.8951} & \textbf{98.298} & 226.1452 & \textbf{0.2178} \\
\multirow{-2}{*}{\cellcolor[HTML]{D3D3D3}\textbf{+sr}} & \cellcolor[HTML]{F0F0F0}Baseline+ & \textbf{25.2008} & \textbf{0.6887} & 5.8415 & 0.3758 & 0.0149 & 141.7857 & 140.7089 & 98.082 & \textbf{225.0139} & 0.2195 \\
\hline
\cellcolor[HTML]{D3D3D3} & \cellcolor[HTML]{F0F0F0}PS & 27.6859 & 0.7906 & 6.6324 & 0.2998 & 0.0121 & 50.6460 & 50.3979 & \textbf{103.968} & 133.2403 & 0.1652 \\
\multirow{-2}{*}{\cellcolor[HTML]{D3D3D3}\textbf{deblur}} & \cellcolor[HTML]{F0F0F0}Baseline+ & \textbf{27.7020} & \textbf{0.7913} & \textbf{6.6048} & \textbf{0.2996} & \textbf{0.0117} & \textbf{46.7897} & \textbf{46.8866} & 103.932 & \textbf{132.4128} & \textbf{0.1651} \\
\hline
\cellcolor[HTML]{D3D3D3} & \cellcolor[HTML]{F0F0F0}PS & 25.1441 & 0.6730 & 6.2271 & 0.4246 & 0.0157 & \textbf{147.1632} & \textbf{147.1276} & \textbf{98.856} & 230.0379 & 0.2461 \\
\multirow{-2}{*}{\cellcolor[HTML]{D3D3D3}\textbf{+deblur}} & \cellcolor[HTML]{F0F0F0}Baseline+ & \textbf{25.1720} & \textbf{0.6741} & \textbf{6.2111} & \textbf{0.4235} & \textbf{0.0153} & 147.5224 & 147.4916 & 98.658 & \textbf{228.1795} & \textbf{0.2460} \\
\hline
\cellcolor[HTML]{D3D3D3} & \cellcolor[HTML]{F0F0F0}PS & 26.6410 & 0.7321 & 3.6202 & 0.1638 & 0.0135 & \textbf{195.7243} & \textbf{194.9824} & \textbf{98.514} & 173.3719 & 0.0953 \\
\multirow{-2}{*}{\cellcolor[HTML]{D3D3D3}\textbf{temporal blur}} & \cellcolor[HTML]{F0F0F0}Baseline+ & \textbf{26.6432} & \textbf{0.7323} & \textbf{3.6108} & \textbf{0.1635} & \textbf{0.0133} & 207.6850 & 207.3517 & 98.424 & \textbf{173.1191} & \textbf{0.0952} \\
\hline
\end{tabular}%
}
\end{table*}

\begin{figure}[htbp]
	\centering
	\includegraphics[width=\columnwidth]{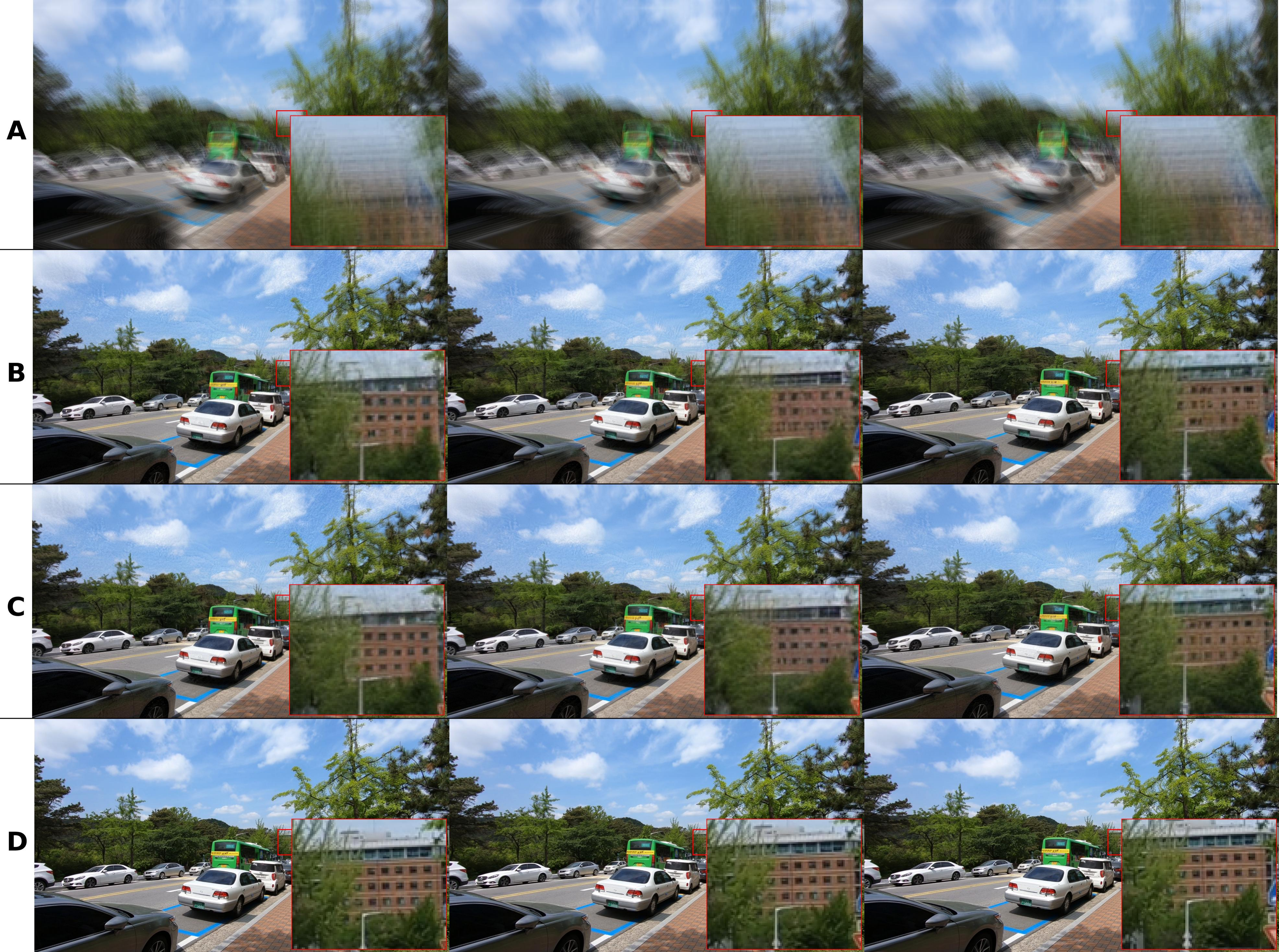}
	\caption{\small Visual comparison of PSG vs. Baseline+ on three successive frames on a sequence from \textbf{REDS} dataset for the Temporal\_blur task: (A) Measurement, (B) Baseline+, (C) PSG, (D) Ground-Truth.}
	\label{fig:PS_temporal_blur}
\end{figure}

\begin{figure}[htbp]
	\centering
	\includegraphics[width=\columnwidth]{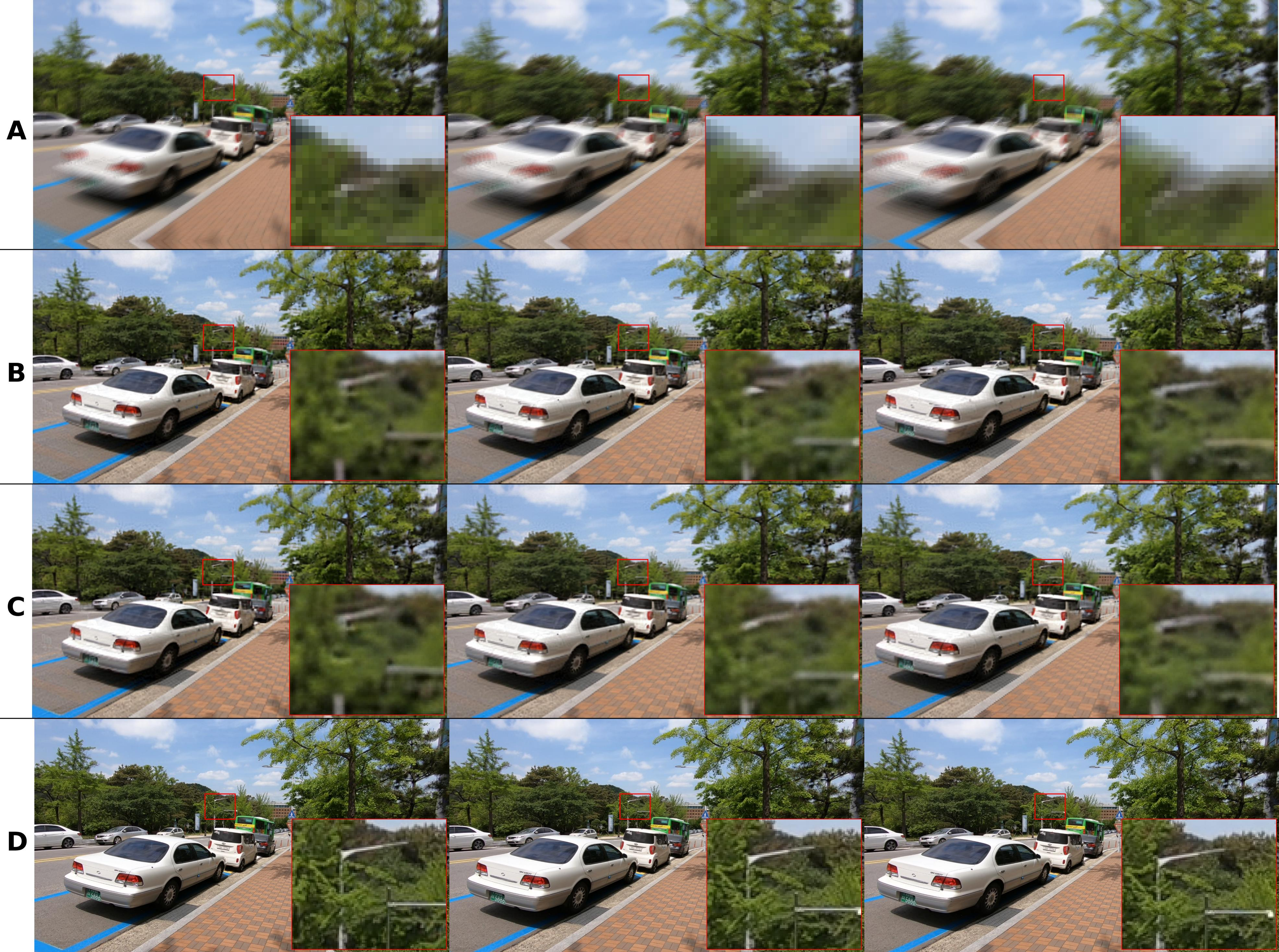}
	\caption{\small Visual comparison of PSG vs. Baseline+ on three successive frames on a sequence from \textbf{REDS} dataset for the +SR task: (A)~Measurement, (B) Baseline+, (C) PSG, (D) Ground-Truth.}
	\label{fig:PS_sr_plus}
\end{figure}

Since perceptual straightening acts as a temporal regularizer, we expect it to leave frame appearance largely unchanged while making motion look more natural by small edits. Hence, we demonstrate the visual effect of the PSG on three consecutive frames of a video to observe motion continuity. As Figures \ref{fig:PS_temporal_blur} and  \ref{fig:PS_sr_plus} depicts,  perceptual straightening guidance on the temporal blur and SR+ tasks leads to reduced micro-wobble in structures like building edges, window frames, less boiling of repetitive textures and cleaner line persistence.

\subsection{Effect of Multi-Path Ensemble Sampling (MPES)}
\label{subsec:ablation-multi-path-ensemble}
To evaluate the effectiveness of our MPES framework introduced in Section \ref{subsec: ensemble_stratefies}, we performed  ablation studies to evaluate key aspects of MPES: (1)  the ideal fusion representation space (latent vs pixel) and (2) the effect of ensemble size (K=2 vs K=3).

\setcounter{paragraph}{0}

\paragraph{Experiment 1} Latent-Space Fusion (K=2)

In this setup, two independent samplings are run in parallel  during the denoising. At the final step, the predicted latent representations are fused just before decoding. A refinement step is used after fusion. 

\paragraph{Experiment 2}  Pixel-Space Fusion (K=2) or (K=3)

In this case, the fusion is applied in the pixel space after decoding K=2  or K=3 samples. A refinement step is again used after fusion. 
The experiment with K=3 investigates whether increasing the number of trajectories leads to improved robustness based on the intuition that multiple diverse reconstructions can reduce artifacts through aggregation.

\begin{figure}[H]
	\centering
	\includegraphics[width=\columnwidth]{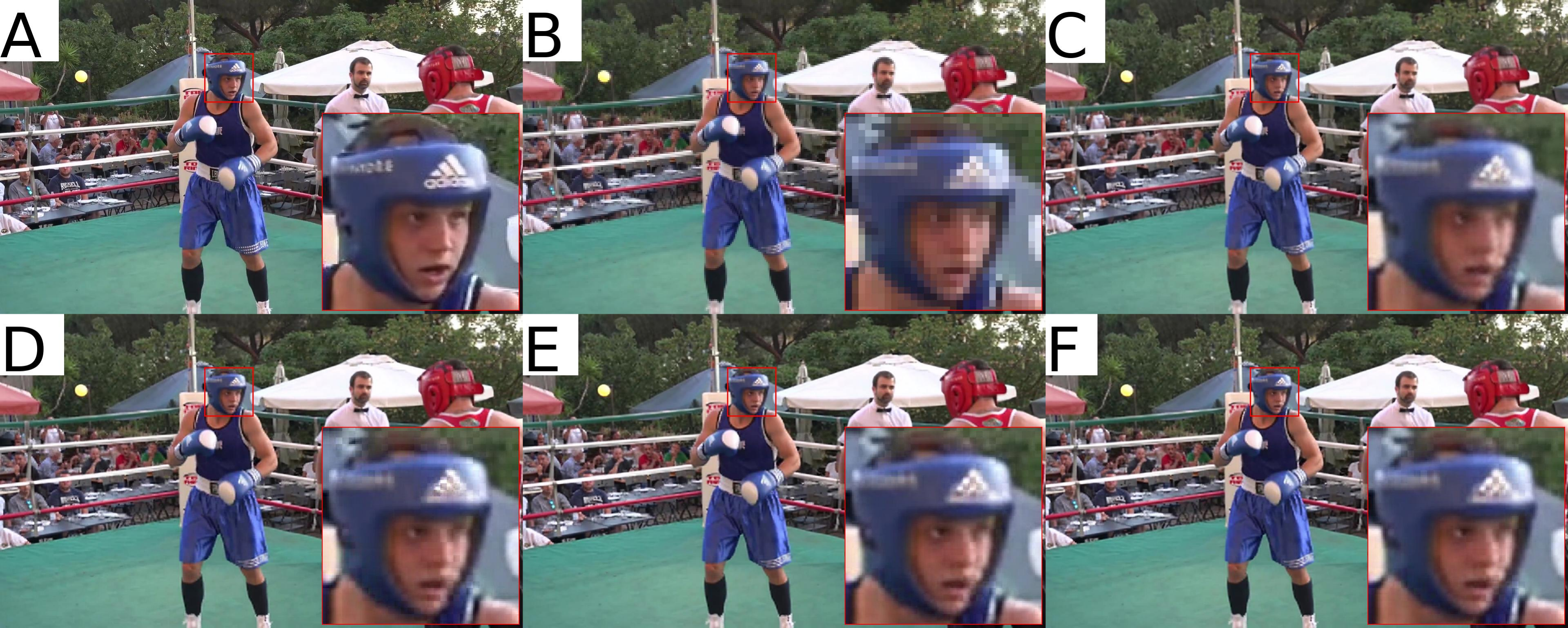}
	\caption{\small Visual comparison of \textbf{Ensemble sampling} vs. Baseline+ for the \textbf{SR} task on \textbf{DAVIS} dataset: (A) Ground-Truth, (B) Measurement, (C) Baseline+, (D) Latent Fusion, K=2, (E) Pixel Fusion, K=2, (F)~Pixel Fusion, K=3.}
	\label{fig:ensemble_sr}
\end{figure}
\vspace{-5pt}

\begin{figure}[H]
	\centering
	\includegraphics[width=\columnwidth]{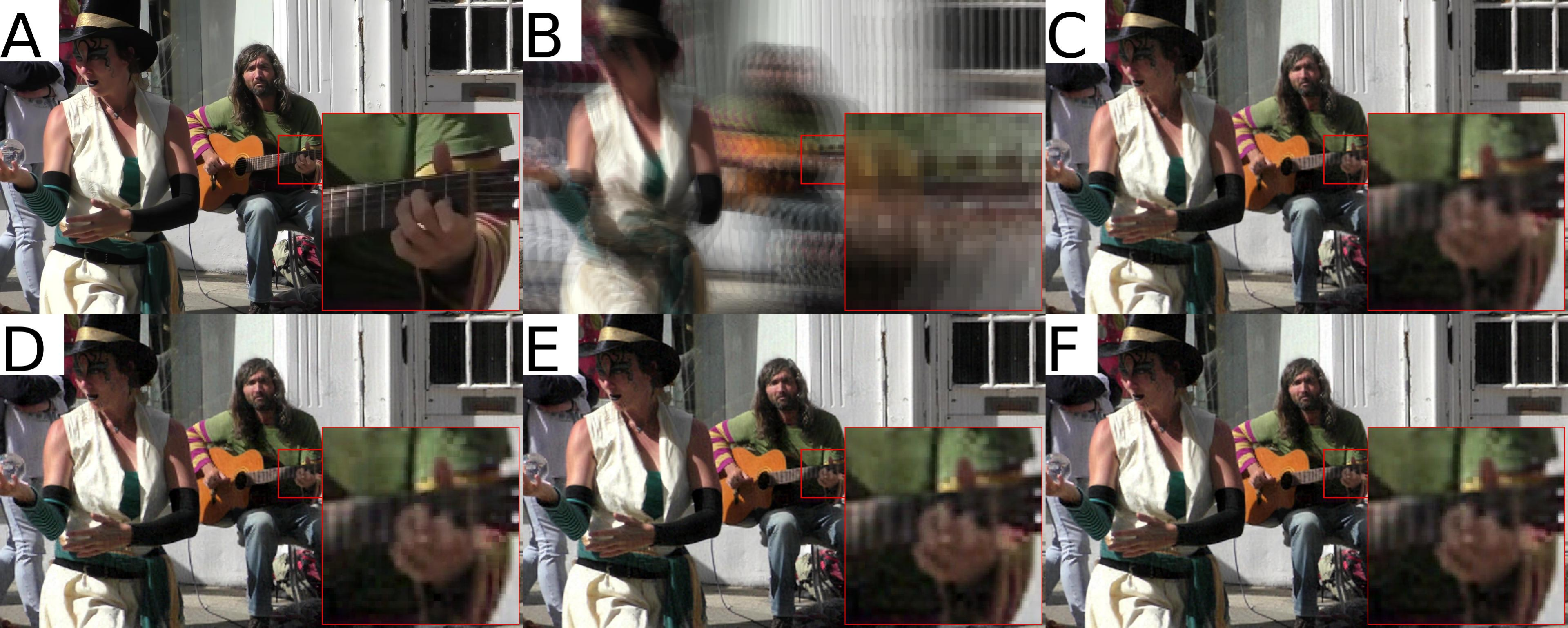}
	\caption{\small Visual comparison of \textbf{Ensemble sampling} vs. Baseline+ for the \textbf{+SR} task on \textbf{DAVIS} dataset:  (A) Ground-Truth, (B) Measurement, (C) Baseline+, (D) Latent Fusion, K=2, (E) Pixel Fusion, K=2, (F)~Pixel Fusion, K=3.}
	\label{fig:ensemble_+sr}
\end{figure}
\vspace{-5pt}

\begin{figure}[H]
	\centering
	\includegraphics[width=\columnwidth]{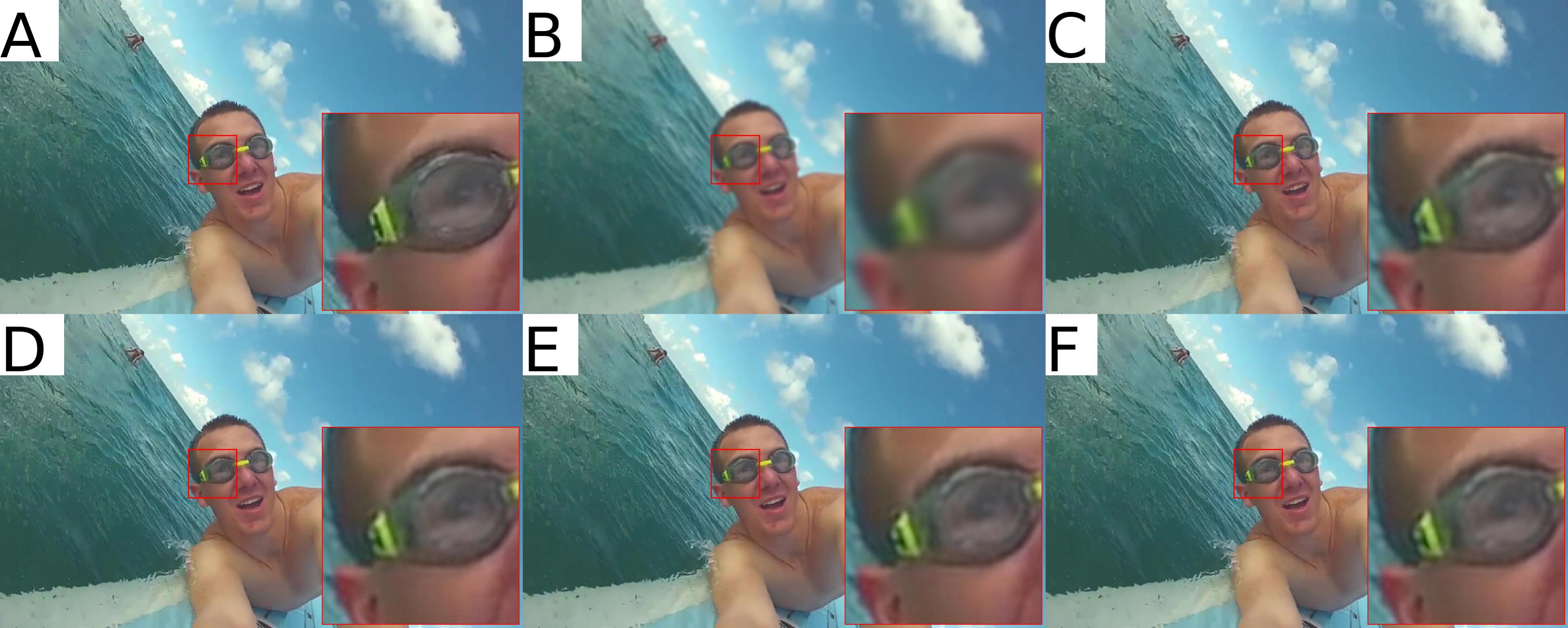}
	\caption{\small Visual comparison of \textbf{Ensemble sampling} vs. Baseline+  for the \textbf{Deblur} task on \textbf{DAVIS} dataset: (A) Ground-Truth, (B) Measurement, (C) Baseline+, (D) Latent Fusion, K=2, (E) Pixel Fusion, K=2, (F)~Pixel Fusion, K=3.}
	\label{fig:ensemble_deblur}
\end{figure}
\vspace{-5pt}

\begin{figure}[H]
	\centering
	\includegraphics[width=\columnwidth]{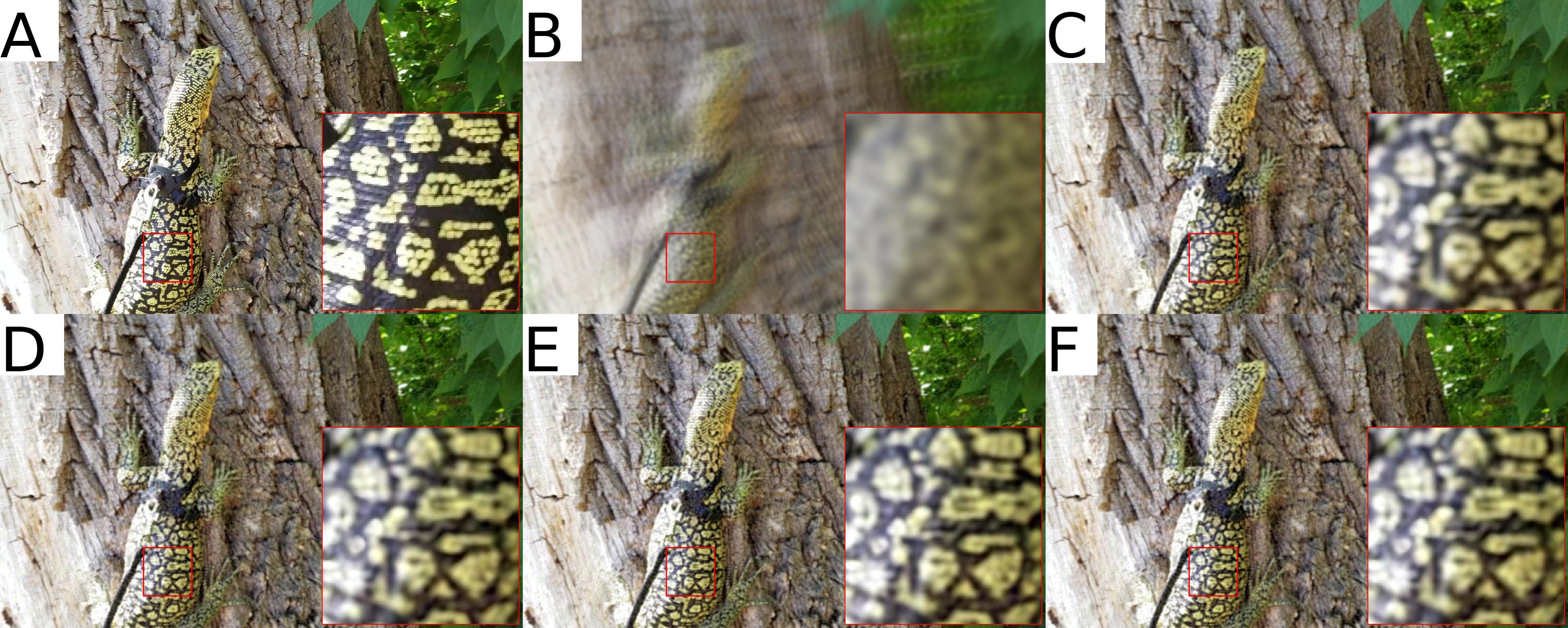}
	\caption{\small Visual comparison of \textbf{Ensemble sampling} vs. Baseline+ for the \textbf{+Deblur} task on \textbf{DAVIS} dataset: (A) Ground-Truth, (B) Measurement, (C) Baseline+, (D) Latent Fusion, K=2, (E) Pixel Fusion, K=2, (F)~Pixel Fusion, K=3.}
	\label{fig:ensemble_+deblur}
\end{figure}
\vspace{-5pt}

\begin{figure}[H]
	\centering
	\includegraphics[width=\columnwidth]{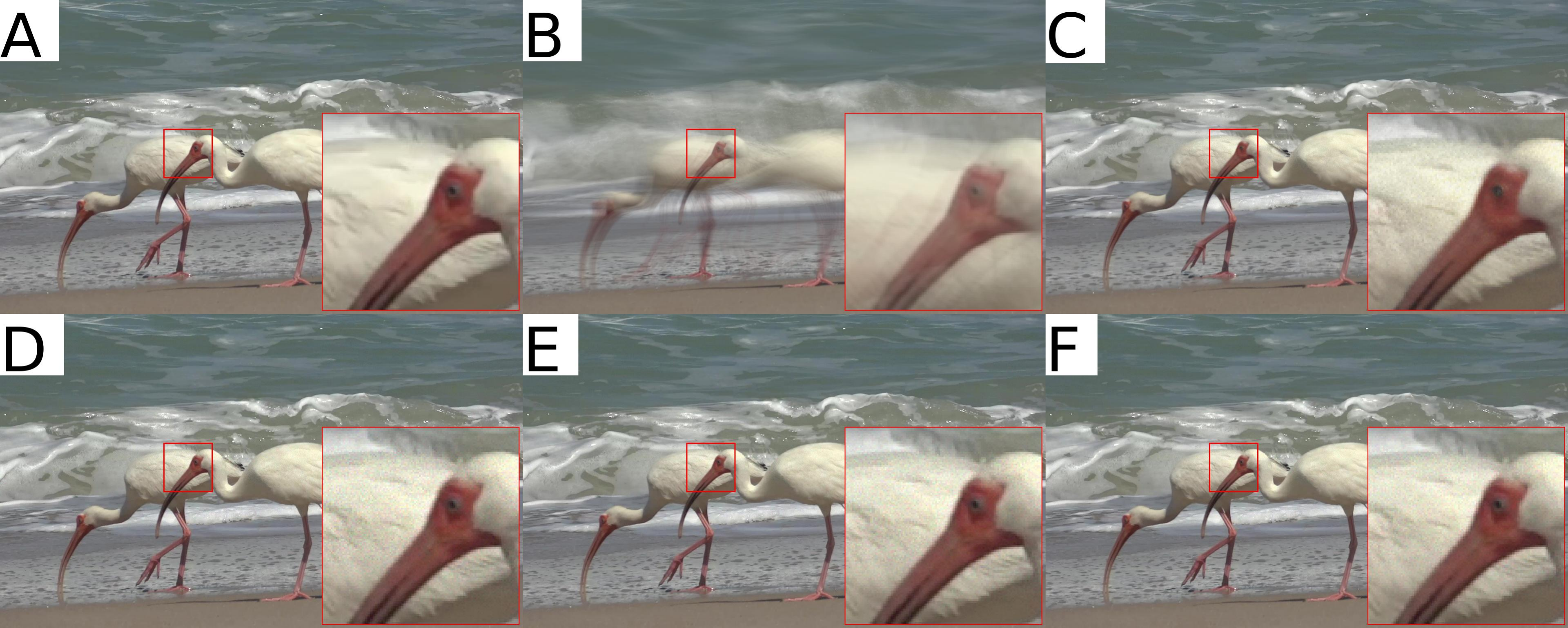}
	\caption{\small Visual comparison of \textbf{Ensemble sampling} vs. Baseline+ for the \textbf{Temporal Blur} task on \textbf{DAVIS} dataset: (A) Ground-Truth, (B) Measurement, (C) Baseline+, (D) Latent Fusion, K=2, (E) Pixel Fusion, K=2, (F)~Pixel Fusion, K=3.}
	\label{fig:ensemble_temporal}
\end{figure}
\vspace{-5pt}

\begin{table*}[htbp]
\centering
\caption{Quantitative evaluation on \textbf{DAVIS} dataset comparing different \textbf{Ensemble} ablation experiments across various degradation tasks. Best values are \textbf{bolded} and \textcolor{blue}{blue} values represent the second-best scores}
\label{tab:ensemble_davis}
\resizebox{\textwidth}{!}{%
\begin{tabular}{|c|c|cccc|cccc|cc|}
\hline
& & \multicolumn{4}{c|}{\textbf{Spatial}} & \multicolumn{4}{c|}{\textbf{Temporal}} & \multicolumn{2}{c|}{\textbf{Spatio-Temporal}} \\
\cline{3-12}
\textbf{Degradation} & \textbf{Method} & \textbf{PSNR↑} & \textbf{SSIM↑} & \textbf{NIQE↓} & \textbf{LPIPS↓} & \textbf{flow\_MSE↓} &  \textbf{fvd-videogpt↓} & \textbf{fvd-styleganv↓} & \textbf{Straightness↑} & \textbf{D-ST↓} & \textbf{P-ST↓} \\
\hline

\cellcolor[HTML]{D3D3D3} & Latent Fusion (K=2) & 30.2939 & 0.8520 & \textbf{5.6751} & \textbf{0.1963} & 0.0137 & 70.5963 & 70.3234 & 96.048 & 146.6228 & \textbf{0.1171} \\
\cellcolor[HTML]{D3D3D3} & Pixel Fusion, (K=2) & {\color[HTML]{0000FF} 30.4857} & {\color[HTML]{0000FF} 0.8558} & {\color[HTML]{0000FF} 5.7019} & {\color[HTML]{0000FF} 0.1988} & {\color[HTML]{0000FF} 0.0134} & \textbf{67.5485} & \textbf{67.6531} & {\color[HTML]{0000FF} 96.084} & {\color[HTML]{0000FF} 144.3505} & {\color[HTML]{0000FF} 0.1185} \\
\cellcolor[HTML]{D3D3D3} & Pixel Fusion, (K=3) & \textbf{30.5181} & \textbf{0.8566} & 5.7677 & 0.2011 & \textbf{0.0134} & {\color[HTML]{0000FF} 69.4153} & {\color[HTML]{0000FF} 68.6170} & \textbf{96.086} & \textbf{143.7916} & 0.1199 \\
\multirow{-5}{*}{\cellcolor[HTML]{D3D3D3}\textbf{sr}} & Baseline+ & 29.7367 & 0.8294 & 6.1289 & 0.2249 & 0.0145 & 83.8501 & 82.2478 & 95.454 & 159.4358 & 0.1350 \\
\hline

\cellcolor[HTML]{D3D3D3} & Latent Fusion (K=2) & 28.9560 & 0.8003 & \textbf{5.4961} & \textbf{0.2633} & 0.0148 & 99.8820 & 98.6967 & 92.250 & 165.5139 & {\color[HTML]{0000FF} 0.1635} \\
\cellcolor[HTML]{D3D3D3} & Pixel Fusion, (K=2) & {\color[HTML]{0000FF} 29.2106} & {\color[HTML]{0000FF} 0.8076} & {\color[HTML]{0000FF} 5.6564} & 0.2659 & {\color[HTML]{0000FF} 0.0144} & {\color[HTML]{0000FF} 92.6848} & {\color[HTML]{0000FF} 91.0386} & {\color[HTML]{0000FF} 92.844} & {\color[HTML]{0000FF} 160.9432} & 0.1641 \\
\cellcolor[HTML]{D3D3D3} & Pixel Fusion, (K=3) & \textbf{29.2684} & \textbf{0.8114} & 5.7581 & {\color[HTML]{0000FF} 0.2643} & \textbf{0.0142} & \textbf{91.3272} & \textbf{90.1016} & \textbf{92.952} & \textbf{159.7503} & \textbf{0.1629} \\
\multirow{-5}{*}{\cellcolor[HTML]{D3D3D3}\textbf{+sr}} & Baseline+ & 28.5687 & 0.7878 & 6.2499 & 0.2869 & 0.0151 & 122.4390 & 120.0357 & 92.322 & 178.1407 & 0.1781 \\
\hline

\cellcolor[HTML]{D3D3D3} & Latent Fusion (K=2) & 32.4332 & 0.8780 & \textbf{6.1444} & {\color[HTML]{0000FF} 0.2196} & \textbf{0.0108} & 64.6716 & 64.7970 & 96.390 & 107.5168 & 0.1305 \\
\cellcolor[HTML]{D3D3D3} & Pixel Fusion, (K=2) & {\color[HTML]{0000FF} 32.5858} & {\color[HTML]{0000FF} 0.8807} & {\color[HTML]{0000FF} 6.1676} & 0.2196 & 0.0111 & \textbf{63.7341} & \textbf{64.3902} & {\color[HTML]{0000FF} 96.390} & {\color[HTML]{0000FF} 106.0960} & {\color[HTML]{0000FF} 0.1305} \\
\cellcolor[HTML]{D3D3D3} & Pixel Fusion, (K=3) & \textbf{32.6534} & \textbf{0.8834} & 6.1907 & \textbf{0.2163} & {\color[HTML]{0000FF} 0.0111} & {\color[HTML]{0000FF} 64.4063} & {\color[HTML]{0000FF} 64.5867} & \textbf{96.392} & \textbf{105.7445} & \textbf{0.1286} \\
\multirow{-5}{*}{\cellcolor[HTML]{D3D3D3}\textbf{deblur}} & Baseline+ & 31.1342 & 0.8529 & 6.5927 & 0.2302 & 0.0123 & 69.1178 & 68.6272 & 95.616 & 130.3061 & 0.1379 \\
\hline

\cellcolor[HTML]{D3D3D3} & Latent Fusion (K=2) & 28.5169 & {\color[HTML]{0000FF} 0.7770} & \textbf{6.3212} & {\color[HTML]{0000FF} 0.3312} & 0.0150 & 147.8147 & 146.2795 & 92.340 & 174.9943 & {\color[HTML]{0000FF} 0.2055} \\
\cellcolor[HTML]{D3D3D3} & Pixel Fusion, (K=2) & {\color[HTML]{0000FF} 28.7287} & 0.7768 & {\color[HTML]{0000FF} 6.3384} & 0.3358 & \textbf{0.0146} & {\color[HTML]{0000FF} 142.6238} & {\color[HTML]{0000FF} 139.6994} & {\color[HTML]{0000FF} 93.150} & {\color[HTML]{0000FF} 168.7994} & 0.2065 \\
\cellcolor[HTML]{D3D3D3} & Pixel Fusion, (K=3) & \textbf{28.7903} & \textbf{0.7818} & 6.3522 & \textbf{0.3309} & {\color[HTML]{0000FF} 0.0149} & \textbf{142.5426} & \textbf{139.6423} & \textbf{93.222} & \textbf{168.2588} & \textbf{0.2034} \\
\multirow{-5}{*}{\cellcolor[HTML]{D3D3D3}\textbf{+deblur}} & Baseline+ & 28.1199 & 0.7735 & 6.6476 & 0.3473 & 0.0159 & 169.0877 & 167.1298 & 92.250 & 189.5531 & 0.2157 \\
\hline

\cellcolor[HTML]{D3D3D3} & Latent Fusion (K=2) & 33.5976 & 0.8168 & 5.4338 & 0.1772 & 0.0115 & 17.2757 & 16.9627 & 93.564 & 44.3696 & 0.1085 \\
\cellcolor[HTML]{D3D3D3} & Pixel Fusion, (K=2) & {\color[HTML]{0000FF} 34.7168} & {\color[HTML]{0000FF} 0.8469} & 5.2299 & {\color[HTML]{0000FF} 0.1481} & \textbf{0.0109} & {\color[HTML]{0000FF} 12.0802} & {\color[HTML]{0000FF} 12.5348} & 93.978 & {\color[HTML]{0000FF} 36.4243} & {\color[HTML]{0000FF} 0.0903} \\
\cellcolor[HTML]{D3D3D3} & Pixel Fusion, (K=3) & \textbf{35.1709} & \textbf{0.8583} & {\color[HTML]{0000FF} 5.1405} & \textbf{0.1369} & {\color[HTML]{0000FF} 0.0111} & \textbf{10.4322} & \textbf{10.6899} & {\color[HTML]{0000FF} 94.068} & \textbf{34.0602} & \textbf{0.0834} \\
\multirow{-5}{*}{\cellcolor[HTML]{D3D3D3}\textbf{temporal blur}} & Baseline+ & 30.4446 & 0.7938 & \textbf{4.5720} & 0.2149 & 0.0127 & 51.4089 & 50.0101 & \textbf{94.194} & 159.1931 & 0.1308 \\
\hline
\end{tabular}%
}
\end{table*}

\begin{table*}[htbp]
\centering
\caption{Quantitative evaluation on \textbf{REDS4} dataset comparing different \textbf{Ensemble} ablation experiments across various degradation tasks. Best values are \textbf{bolded}.}
\label{tab:ensemble_reds4}
\resizebox{\textwidth}{!}{%
\begin{tabular}{|c|c|cccc|cccc|cc|}
\hline
& & \multicolumn{4}{c|}{\textbf{Spatial}} & \multicolumn{4}{c|}{\textbf{Temporal}} & \multicolumn{2}{c|}{\textbf{Spatio-Temporal}} \\
\cline{3-12}
\textbf{Degradation} & \textbf{Method} & \textbf{PSNR↑} & \textbf{SSIM↑} & \textbf{NIQE↓} & \textbf{LPIPS↓} & \textbf{flow\_MSE↓} & \textbf{fvd-videogpt↓} & \textbf{fvd-styleganv↓} & \textbf{Straightness↑} & \textbf{D-ST↓} & \textbf{P-ST↓} \\
\hline
\cellcolor[HTML]{D3D3D3} & Latent Fusion (K=2) & 26.6506 & 0.7697 & {\color[HTML]{0000FF}5.0066} & \textbf{0.2774} & 0.0126 & 55.8290 & 54.7760 & 104.400 & 162.5172 & \textbf{0.1523} \\
\cellcolor[HTML]{D3D3D3} & Pixel Fusion, (K=2) & {\color[HTML]{0000FF} 26.7709} & {\color[HTML]{0000FF} 0.7747} & \textbf{4.9310} & {\color[HTML]{0000FF} 0.2839 } & {\color[HTML]{0000FF} 0.0124} & {\color[HTML]{0000FF} 50.7844} & {\color[HTML]{0000FF} 51.1504} & {\color[HTML]{0000FF} 104.454} & {\color[HTML]{0000FF} 158.5174} & {\color[HTML]{0000FF} 0.1557} \\
\cellcolor[HTML]{D3D3D3} & Pixel Fusion, (K=3) & \textbf{26.7912} & \textbf{0.7756} & 5.0098 & 0.2870 & \textbf{0.0124} & \textbf{50.1927} & \textbf{50.8866} & \textbf{104.472} & \textbf{157.7888} & 0.1574 \\
\multirow{-5}{*}{\cellcolor[HTML]{D3D3D3}\textbf{sr}} & Baseline+ & 26.2763 & 0.7431 & 5.3510 & 0.3195 & 0.0135 & 51.6672 & 51.8888 & 103.626 & 177.7404 & 0.1767 \\
\hline
\cellcolor[HTML]{D3D3D3} & Latent Fusion (K=2) & 25.5295 & 0.7063 & \textbf{5.3644} & \textbf{0.3366} & {\color[HTML]{0000FF} 0.0143} & 135.6013 & 135.3260 & 98.424 & 208.2861 & \textbf{0.1960} \\
\cellcolor[HTML]{D3D3D3} & Pixel Fusion, (K=2) & {\color[HTML]{0000FF} 25.7114} & {\color[HTML]{0000FF} 0.7158} & {\color[HTML]{0000FF} 5.3732} & {\color[HTML]{0000FF} 0.3421}  & 0.0145 & \textbf{123.7890} & \textbf{123.7817} & {\color[HTML]{0000FF} 98.910} & \textbf{200.6943} & {\color[HTML]{0000FF} 0.1982} \\
\cellcolor[HTML]{D3D3D3} & Pixel Fusion, (K=3) & \textbf{25.7314} & \textbf{0.7175} & 5.4540 & 0.3450 & \textbf{0.0143} & {\color[HTML]{0000FF} 124.5278} & {\color[HTML]{0000FF} 124.0970} & \textbf{98.964} & {\color[HTML]{0000FF} 199.7032} & 0.1997\\
\multirow{-5}{*}{\cellcolor[HTML]{D3D3D3}\textbf{+sr}} & Baseline+ & 25.2008 & 0.6887 & 5.8415 & 0.3758 & 0.0149 & 141.7857 & 140.7089 & 98.082 & 225.0139 & 0.2195 \\
\hline
\cellcolor[HTML]{D3D3D3} & Latent Fusion (K=2) & 28.7609 & 0.8308 & 5.9514 & 0.2912 & 0.0112 & 54.5014 & 52.9646 & 105.029 & 104.6369 & 0.1588 \\
\cellcolor[HTML]{D3D3D3} & Pixel Fusion, (K=2) & {\color[HTML]{0000FF} 28.9183} & {\color[HTML]{0000FF} 0.8361} & \textbf{5.9123} & \textbf{0.2880} & {\color[HTML]{0000FF} 0.0111} & {\color[HTML]{0000FF} 51.8873} & 53.1965 & {\color[HTML]{0000FF}105.029} & {\color[HTML]{0000FF} 101.4920} & \textbf{0.1571} \\
\cellcolor[HTML]{D3D3D3} & Pixel Fusion, (K=3) & \textbf{28.9350} & \textbf{0.8378} & {\color[HTML]{0000FF} 5.9346} & {\color[HTML]{0000FF} 0.2893} & \textbf{0.0111} & 52.5554 & {\color[HTML]{0000FF} 52.6310} & \textbf{105.030} & \textbf{101.1511} & {\color[HTML]{0000FF} 0.1578} \\
\multirow{-5}{*}{\cellcolor[HTML]{D3D3D3}\textbf{deblur}} & Baseline+ & 27.7020 & 0.7913 & 6.6048 & 0.2996 & 0.0117 & \textbf{46.7897} & \textbf{46.8866} & 103.932 & 132.4128 & 0.1651 \\
\hline
\cellcolor[HTML]{D3D3D3} & Latent Fusion (K=2) & 25.5242 & 0.6810 & {\color[HTML]{0000FF} 6.0540} & 0.4088 & \textbf{0.0142} & 134.2759 & 133.3326 & 98.928 & 209.8171 & 0.2368 \\
\cellcolor[HTML]{D3D3D3} & Pixel Fusion, (K=2) & {\color[HTML]{0000FF} 25.7344} & {\color[HTML]{0000FF} 0.6873} & \textbf{6.0460} & \textbf{0.4039}  & {\color[HTML]{0000FF} 0.0146} & {\color[HTML]{0000FF} 126.7702} & {\color[HTML]{0000FF} 125.8548} & {\color[HTML]{0000FF} 99.414} & {\color[HTML]{0000FF} 200.6415} & \textbf{0.2328} \\
\cellcolor[HTML]{D3D3D3} & Pixel Fusion, (K=3) & \textbf{25.7601} & \textbf{0.6910} & 6.0617 & {\color[HTML]{0000FF} 0.4045} & 0.0148 & \textbf{126.4900} & \textbf{125.6219} & \textbf{99.486} & \textbf{199.9967} & {\color[HTML]{0000FF} 0.2330} \\
\multirow{-5}{*}{\cellcolor[HTML]{D3D3D3}\textbf{+deblur}} & Baseline+ & 25.1720 & 0.6741 & 6.2111 & 0.4235 & 0.0153 & 147.5224 & 147.4916 & 98.658 & 228.1795 & 0.2460 \\
\hline
\cellcolor[HTML]{D3D3D3} & Latent Fusion (K=2) & 30.1218 & 0.8552 & \textbf{2.9064} & 0.1215 & 0.0136 & {\color[HTML]{0000FF} 143.9532} & 143.6348 & 98.640 & 82.7805 & 0.0706 \\
\cellcolor[HTML]{D3D3D3} & Pixel Fusion, (K=2) & 30.5346 & {\color[HTML]{0000FF} 0.8660} & 2.9575 & {\color[HTML]{0000FF} 0.1110} & \textbf{0.0127} & \textbf{138.6359} & \textbf{138.4109} & {\color[HTML]{0000FF} 99.414} & {\color[HTML]{0000FF} 76.1517}& {\color[HTML]{0000FF} 0.0639} \\
\cellcolor[HTML]{D3D3D3} & Pixel Fusion, (K=3) & {\color[HTML]{0000FF} 30.6237} & \textbf{0.8699} & {\color[HTML]{0000FF} 2.9538} & \textbf{0.1085} & {\color[HTML]{0000FF} 0.0128} & 143.9935 & {\color[HTML]{0000FF} 142.2372} & \textbf{99.504} & \textbf{75.0959} & \textbf{0.0625} \\
\multirow{-5}{*}{\cellcolor[HTML]{D3D3D3}\textbf{temporal blur}} & Baseline+ & 26.6432 & 0.7323 & 3.6108 & 0.1635 & 0.0133 & 207.6850 & 207.3517 & 98.424 & 173.1191 & 0.0952 \\
\hline
\end{tabular}%
}
\end{table*}

\noindent\textbf{Results and Analysis}

Tables~\ref{tab:ensemble_davis} and \ref{tab:ensemble_reds4} illustrate the performance comparison of ensemble sampling strategies for DAVIS and REDS4 datasets respectively, displaying consistent patterns and also dataset-specific characteristics that we will go over them. Also, The visual results comparing different ensemble sampling strategies with the baseline across multiple degradation types are presented in Figures \ref{fig:ensemble_sr}–\ref{fig:ensemble_temporal}.

\setcounter{paragraph}{0}

\paragraph{Latent vs Pixel Space Fusion}

Pixel-space fusion is generally better than latent-space fusion over different tasks and datasets. On DAVIS inpainting task, for example, Experiment 3 achieves 32.28 dB PSNR versus 31.17 dB in Experiment 2. This superiority potentially occurs because latent-space fusion loses high-frequency information during the single decoding step of averaged latent representations.

\paragraph{Ensemble Size Effects}

Increasing the number of trajectories from K=2 to K=3 slightly improves the results across both datasets specially for spatial fidelity metrics. DAVIS shows gains of 0.4-0.5 dB PSNR (temporal blur: 34.72→35.17 dB) and REDS4 exhibits similar improvements (30.53→30.62 dB). As discussed in section\ref{subsec: ensemble_theory}, these improvements specially in distortion metrics align with ensemble theory that states averaging independent estimators reduces variance. However, the computational cost increases linearly making K=2 more proper in practice.

\paragraph{Cross-Dataset evaluation}

Performance is consistent for both DAVIS and REDS4, but REDS4 shows 3-4 dB lower PSNR values because of the complexity of motion patterns in it. In general,  pixel-space fusion with K=3 shows better metrics on both datasets.

\paragraph{Perception-Distortion Trade-offs}

\begin{figure}[b!]
	\centering
	\includegraphics[width=\columnwidth]{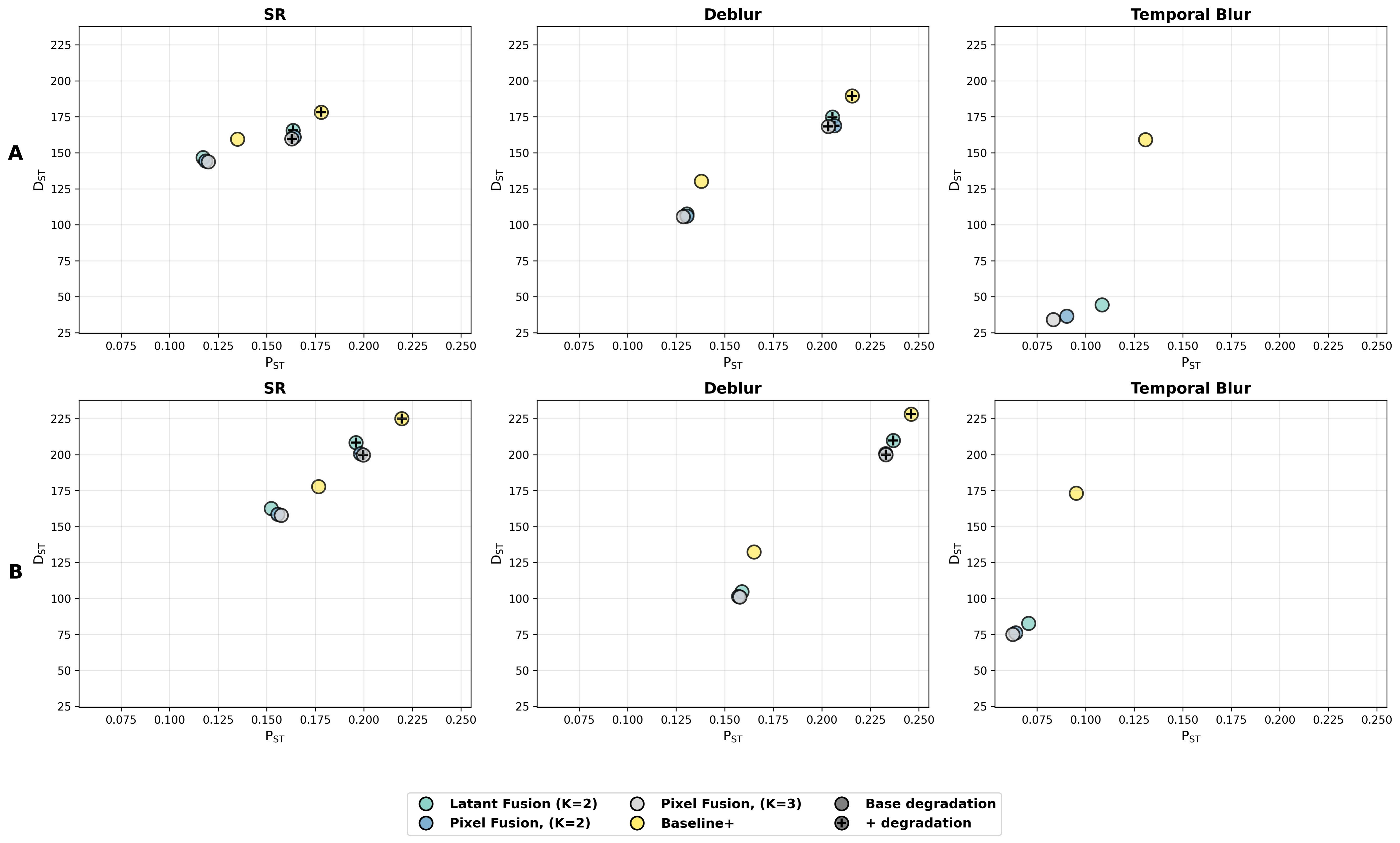} \vspace{-16pt}	

\caption{\small Perception-Distortion trade-off for different Ensemble methods and Baseline+ on (A) \textbf{DAVIS} and (B) \textbf{REDS4} datasets.}
	\label{fig:ensemble_pd_davis}
\end{figure}

To find the method with the best perception-distortion (P-D) balance, Figure~\ref{fig:ensemble_pd_davis}  depicts the perception-distortion trade-off based on $P_{ST}$ (spatio-temporal perceptual quality) and $D_{ST}$ (spatio-temporal distortion) measures proposed in \cite{rahimi2023spatio}.   It can be seen that the proposed multi-path ensembling strategy offers the best P-D trade-off in all cases (closest to the lower left corner).
\vspace{6pt}

\noindent\textbf{Ensemble Method vs. Baseline Comparison}

We compare all three ensembling strategies against the baseline Vision-XL initialized with frame-based DDIM inversion (Baseline+) since the same initialization strategy is used in all Ensemble experiments. Table~\ref{tab:ensemble_davis} and Table~\ref{tab:ensemble_reds4} present comparison results for DAVIS and REDS4 datasets.

\begin{figure*}[t!]
	\centering
	\includegraphics[width=\textwidth]{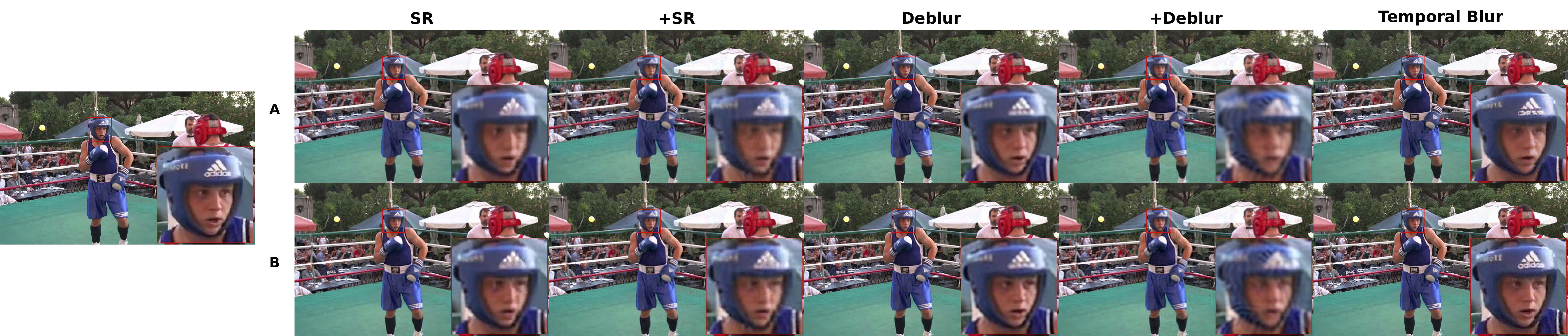}
	\caption{\small Visual comparison \textbf{(A) Baseline+} versus \textbf{(B) pixel fusion (K=2)} on a video from the \textbf{DAVIS} dataset for different degradations.}
	\label{fig:ensemble_k2_baseline+}
\end{figure*}

\setcounter{paragraph}{0}
\paragraph{Universal Ensemble Improvements}
All ensemble methods outperform Baseline+ across almost all restoration tasks and metrics. For example, on DAVIS deblurring,  all independent methods considerably outperform Baseline+: Latent-space K=2 (32.43 dB), Pixel-space K=2 (32.59 dB) and K=3 (32.65 dB) all surpass Baseline+ (31.13 dB) by 1.3-1.5 dB. This consistent improvement indicates that trajectory diversity substantially improves reconstruction quality.

\paragraph{Task-Dependent Performance Improvement} 
By investigating the results, we can see that the benefit of ensemble sampling is correlated  with task difficulty. Simple tasks like SR show moderate 0.5-0.8 dB improvement for PSNR across all ensemble methods, while challenging tasks show huge gains: deblurring improvements range from 1.3-1.5 dB, and temporal degradations achieve 3.2-4.7 dB gains. This observation shows the importance of ensemble diversity as inverse problems approach face limitations with single path.

Figure \ref{fig:ensemble_k2_baseline+} shows the comparison results of pixel-space fusion (k=2) versus Baseline+ for all restoration tasks. As it is shown from the results, ensemble method improved the restoration quality for all tasks.

\paragraph{Ensemble Method Generalizability}
In spite of different content characteristics of DAVIS and REDS4 and different degradation types, ensemble methods show consistent improvement over Baseline+. This observation strongly validate the generalization ability of our ensemble sampling framework.

\paragraph{Computational Efficiency}
Independent pixel-fusion with K=2 provides the best balance between performance (across both datasets) and computational efficiency. It achieves similar performance as Independent pixel-fusion with K=3, while requiring only about double (rather than triple) of the computational load of Baseline+. 

The results indicate that independent pixel fusion ensemble sampling has significant and consistent improvements over single-path Vision-XL baseline. Specifically for challenging inverse problems, where single path sampling may lead to suboptimal solutions, the role of trajectory diversity becomes more crucial and it is worth the additional computation.

\section{Conclusion and Future Directions}

We propose two complementary inference-time strategies: i) Perceptual Straightening Guidance (PSG) and ii) Multi-Path Ensemble Sampling (MPES) to improve temporal consistency and fidelity in zero-shot diffusion-based video restoration.

The proposed PSG leverages a neuroscience-inspired hypothesis by introducing a curvature penalty in the retinal–V1 space. It is used both as a perceptual metric and as a guidance term to refine denoising trajectories. In particular, when degradations contain temporal blur (\texttt{sr+}, \texttt{deblur+}, \texttt{temporal-blur}), the improvement in Straightness is more pronounced and aligns better with FVD. One explanation is that perceptual straightening improves frame-to-frame smoothness, correcting temporal artifacts such as motion smearing and micro-wobble in structures like edges and textures.

The proposed MPES strategy shows that averaging multiple diffusion trajectories leads to consistent gains in both spatial fidelity, perceptual quality, and temporal coherence across diverse restoration tasks. Ensemble methods help reduce stochastic artifacts and stabilize frame evolution without requiring retraining. Despite the theoretical risk of blurring due to MMSE approximation, our analysis shows that correlation among trajectories and concentration near the natural manifold preserves structure and enhances stability.

These findings highlight that simple, training-free strategies like PSG and MPES can significantly improve zero-shot diffusion-based video restoration.
There are several potential directions for future extensions of our work:

\begin{itemize}
    \item \textbf{Better Perceptual Embedders:} Future work can adopt feature spaces that were shown to separate natural and AI-generated video trajectories, such as those used in perceptual-straightening-based detectors~\cite{interno2025ai}, for both the metric and the guidance.

    \item \textbf{Partial-path Guidance:}  Perceptual-straightening guidance can be applied only in the mid/late denoising steps or on shorter frame sequences to avoid over-smoothing.

    \item \textbf{Improved Curvature Penalty:} One can adopt margin-based (hinge/Huber) curvature penalties with adaptive thresholds and a scheduler that increases the PS update weight as samples approach the image manifold.

    \item \textbf{Advanced Fusion Strategies:} Developing adaptive fusion mechanisms, such as attention-based or quality-aware weighting mechanisms, that dynamically combine results based on reconstruction confidence.

    \item \textbf{Generalization Studies:} Testing multi-path ensemble sampling using different diffusion architectures (e.g., Stable Diffusion 1.5~\cite{Rombach_2022_CVPR}, pix-art~\cite{chen2023pixartalpha}, Consistency Models) and broader domains.

    \item \textbf{Partial-Path Ensembling:} Instead of running $K$ full trajectories, we could run $K$ trajectories only for initial denoising steps (e.g., 30--50\% of the denoising process), then fuse them into a single path for the remaining steps. This reduces computational cost while still using early-stage diversity to stabilize reconstruction.
\end{itemize}

\appendix

\subsection{Diffusion Models for Inverse Problems}
\label{subsec:diffusion_inverse_problem}

Denoising Diffusion Probabilistic Models (DDPMs)~\cite{ho2020denoising}, employ a two-step process: a forward diffusion process that adds noise to an image and a learned reverse process that gradually removes the noise.  Latent Diffusion Models (LDMs)~\cite{rombach2022high} perform diffusion in a lower-dimensional latent space $\mathcal{Z}$ to improve the computational efficiency. 

Let  $\mathbf{x}_0 \in \mathbb{R}^{H \times W \times C}$ be the clean image and $(E_\theta, D_\theta)$ be a pretrained VAE encoder–decoder pair, then:
\begin{equation}
\mathbf{z}_0 = E_\theta(\mathbf{x}_0), \quad \hat{\mathbf{x}}_0 = D_\theta(\mathbf{z}_0),
\end{equation}
where $\mathbf{z}_0 \in \mathbb{R}^{h \times w \times c}, \; h \ll H,\; w \ll W$.
The forward process over $T$ timesteps is defined in $\mathcal{Z}$ as:
\begin{equation}
q(\mathbf{z}_{t} | \mathbf{z}_{t-1}) := \mathcal{N}(\mathbf{z}_{t}; \sqrt{1 - \beta_t} \mathbf{z}_{t-1}, \beta_t \mathbf{I}),
\end{equation}
with the closed-form:
\begin{equation}
q(\mathbf{z}_{t} | \mathbf{z}_0) = \mathcal{N}(\mathbf{z}_{t}; \sqrt{\bar{\alpha}_t} \mathbf{z}_0, (1 - \bar{\alpha}_t)\mathbf{I}),
\end{equation}
where $\bar{\alpha}_t := \prod_{s=1}^{t} (1 - \beta_s)$.

The reverse process is modeled as:
\begin{equation}
p_{\theta}(\mathbf{z}_{t-1} | \mathbf{z}_t) := \mathcal{N}(\mathbf{z}_{t-1}; \boldsymbol{\mu}_{\theta}(\mathbf{z}_t, t), \Sigma_{\theta}(\mathbf{z}_t, t)),
\end{equation}
Experimentally, \cite{dhariwal2021diffusion} chose to set $\Sigma_{\theta}(\mathbf{z}_t, t) =\beta_t \mathbf{I}$ while the mean $\boldsymbol{\mu}_{\theta}$ is obtained by predicting the noise $\boldsymbol{\epsilon}$ added in the forward process.  Diffusion models are trained by optimizing the simple mean-squared error (MSE) loss between the true and predicted noise:
\begin{equation}
\mathcal{L}_{\text{simple}} = \mathbb{E}_{\mathbf{z}_0, \boldsymbol{\epsilon}, t} \left[ \left\| \boldsymbol{\epsilon} - \boldsymbol{\epsilon}_{\theta}(\mathbf{z}_t, t) \right\|^2 \right].
\end{equation}

\noindent\textbf{Bayesian Posterior Sampling:}
Restoration is interpreted as sampling from the posterior:
\begin{equation}
p_\theta(\mathbf{x}_0 | \mathbf{y}) \propto p_\theta(\mathbf{x}_0) \cdot p(\mathbf{y} | \mathbf{x}_0),
\end{equation}
where we aim to estimate an unknown signal $\mathbf{x}_0$ from degraded observations $\mathbf{y} = A(\mathbf{x}_0)$, with $A$ being the corruption process, $p_\theta(\mathbf{x}_0)$ is the learned prior provided by pre-trained diffusion model and $p(\mathbf{y} | \mathbf{x}_0)$ is the likelihood eflecting how well a reconstructed image is consistent with observed measurement. In diffusion-based inverse problem solvers the main problem is how to inject the conditioning information from $p(\mathbf{y} | \mathbf{x}_0)$ into this generative process.

To do so, one strategy is guiding the reverse process using intermediate denoised estimates $\hat{\mathbf{z}}_0^{(t)}$, computed via Tweedie's formula \cite{efron2011tweedie}:

\begin{equation}
\label{eq: tweedie's formula}
\hat{\mathbf{z}}_0^{(t)} = \frac{1}{\sqrt{\bar{\alpha}_t}} \left( \mathbf{z}_t - \sqrt{1 - \bar{\alpha}_t} \cdot \boldsymbol{\epsilon}_\theta(\mathbf{z}_t, t) \right),
\end{equation}
and then decoding the estimated clean latent to pixel space and refining it by minimizing a data-consistency loss:
\begin{equation}
\mathcal{L}_{\text{data}}(\hat{\mathbf{x}}_0^{(t)}) = \| A(\hat{\mathbf{x}}_0^{(t)}) - \mathbf{y} \|^2.
\end{equation}
The refined sample is re-noised and the reverse process continues. This iterative refinement effectively integrates measurement constraints into the generative framework, making diffusion models suitable for a wide range of inverse problems.

\subsection{Perceptual Representation in PSH}
\label{apx: PSH}

\begin{figure}[h]
	\centering
	\includegraphics[width=\columnwidth]{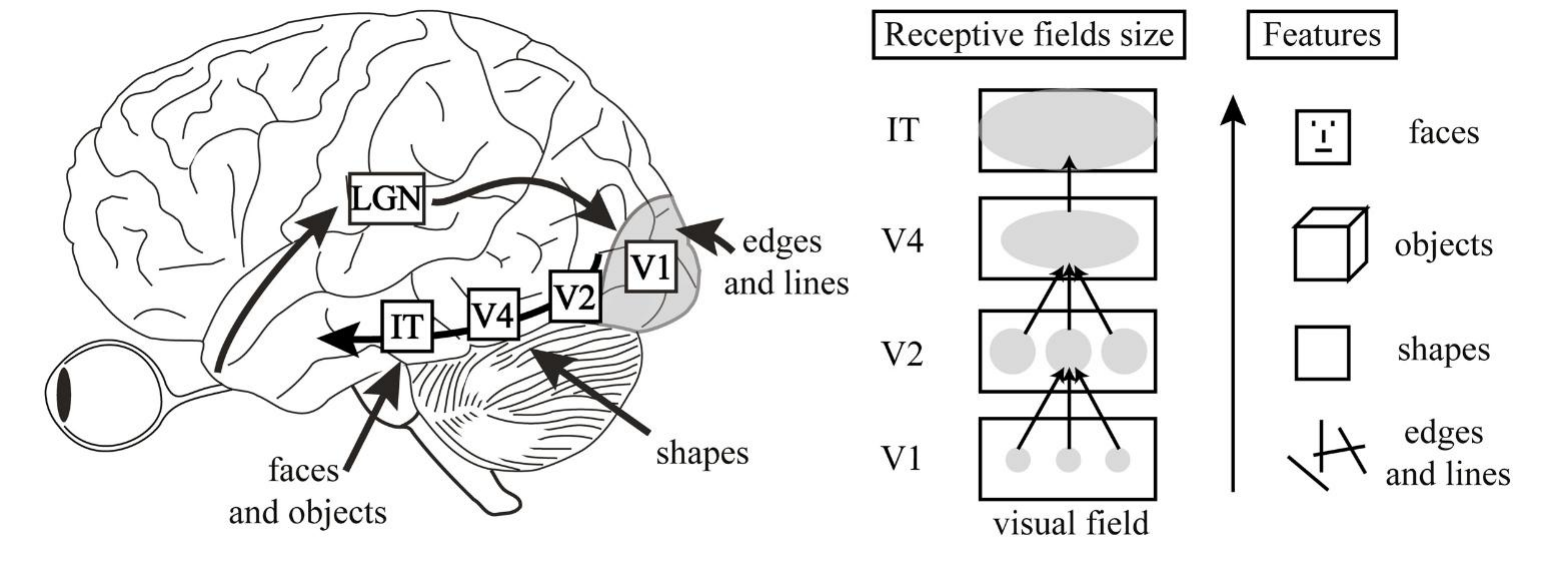}    \vspace{-20pt}
    
	\caption{\small Early processing of human visual system \cite{herzog2014vision}}
	\label{fig:brain}
\end{figure}

\begin{figure}[h!]
	\centering
	\includegraphics[width=\columnwidth]{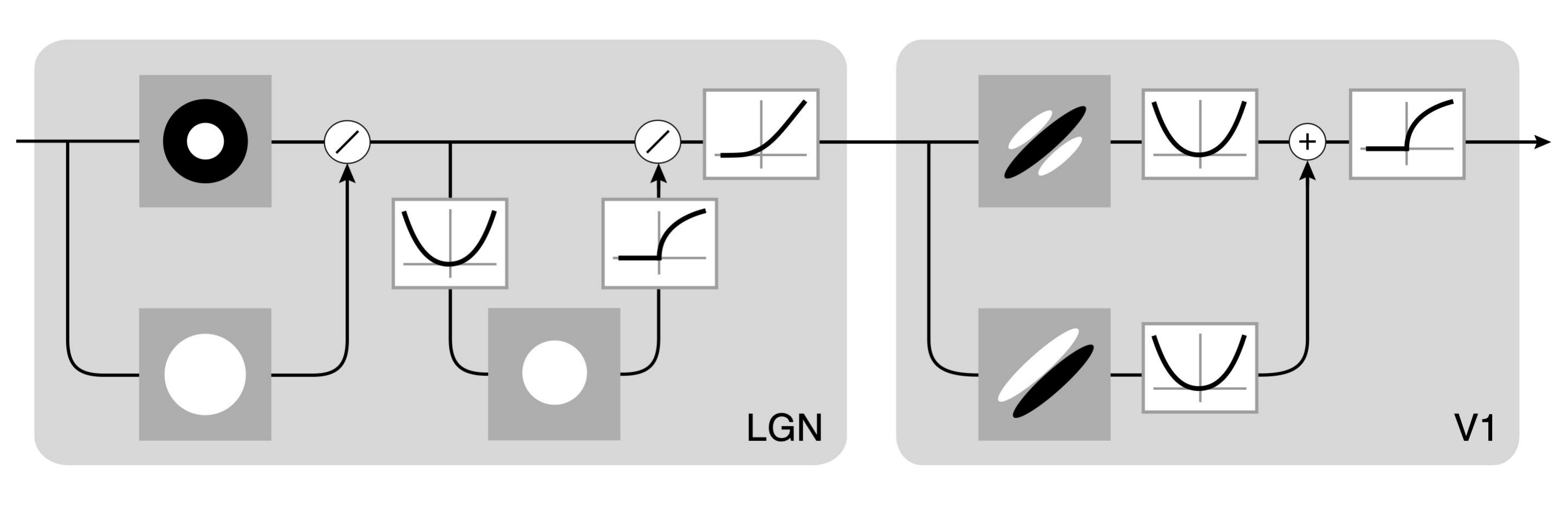} \vspace{-22pt}  
    
    \caption{\small  Two-stage cascade model describing computations in the~retina, lateral geniculate nucleus (LGN) and V1~\cite{henaff2019perceptual}}
	\label{fig:RetinaND+V1}
\end{figure}

The perceptual representation for a video is computed through a two-stage nonlinear model suggested by~\cite{henaff2019perceptual} that simulates the early processing of human visual system illustrated in Figure \ref{fig:brain}.  The two-stage perceptual encoder structure is illustrated in Figure \ref{fig:RetinaND+V1}.

\paragraph{Stage1: RetinaND}

\textit{RetinaND} module is designed to mimic the early visual transformations in the retina and lateral geniculate nucleus (LGN) of the human's visual system.  It converts the video input into a form that is  more aligned with human vision. It normalizes the lighting conditions, emphasizes edges and motion transitions to provide a robust representation for subsequent spatiotemporal modeling. 
The~RetinaND module, implemented in the official codebase~\cite{henaff2019perceptual}, consists of three convolutional layers,  corresponding to:

\begin{enumerate}
    \item \textbf{Center-surround filtering}: enhances local contrast and edges via a Difference-of-Gaussians kernel, implemented as a linear convolution:
    \begin{equation}
        y = k_{\text{lin}} * x
    \end{equation}
    where $x$ is the input frame, $k_{\text{lin}}$ is the center-surround filter, and $*$ shows 2D convolution.

    \item \textbf{Luminance gain control}: normalizes the filtered response using a smoothed estimate of local brightness:
    \begin{equation}
        l = k_{\text{lum}} * x, \qquad y \leftarrow \frac{y}{1 + \alpha \cdot l}
    \end{equation}
    where $k_{\text{lum}}$ is the luminance smoothing kernel, and $\alpha$ is a fixed scalar gain parameter.

    \item \textbf{Contrast gain control}: normalizes the signal based on a local measure of contrast energy:
    \begin{equation}
        c = \sqrt{k_{\text{con}} * y^2 + \varepsilon}, \qquad y \leftarrow \frac{y}{1 + \beta \cdot c}
    \end{equation}
    where $k_{\text{con}}$ is a contrast kernel, $\beta$ is another gain parameter, and $\varepsilon$ is a small constant for numerical stability.

    \item \textbf{Nonlinear activation}, applies Softplus nonlinearity to enhance robustness and biological plausibility:
    \begin{equation}
        y \leftarrow \log\left(1 + e^y\right)
    \end{equation}
\end{enumerate}

\begin{figure}[t]
	\centering
	\includegraphics[width=\columnwidth]{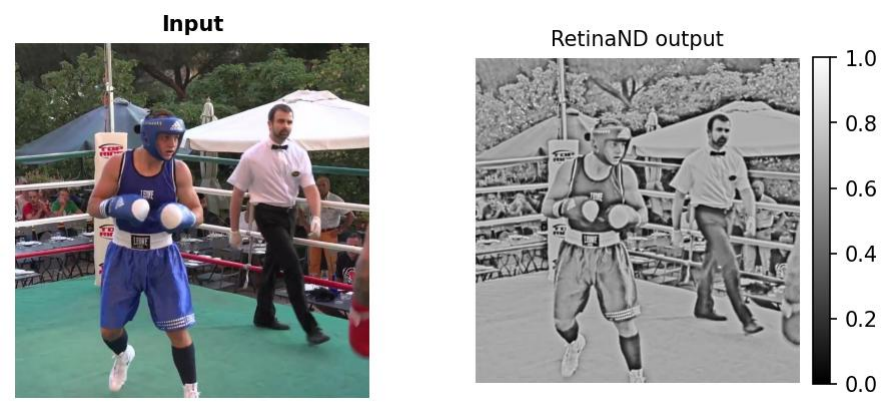}
  \vspace{-18pt}  
  
  \caption{\small Output of RetinaND module}
	\label{fig:RetinaND}
\end{figure}

All filters and normalization constants are designed analytically using known biological response functions.
The input and output of RetinaND module for a sample image are depicted in Figure \ref{fig:RetinaND}.

\paragraph{Stage 2: V1 Feature Extraction}

This stage of the~perceptual model mimics the primary visual cortex~(V1) whose responses are known to be tuned to specific spatial orientations and scales. A \textit{Steerable Pyramid} decomposition is used to extract orientation-, scale-, and frequency-selective responses from the output of the RetinaND module.

The Steerable Pyramid is a linear, multiscale, and orientation-selective transform that uses a set of separable bandpass and lowpass filters in the Fourier domain to  extract multi-scale, orientation-selective features through the  steps:

\begin{enumerate}
    \item The input $y$ is transformed to the frequency domain:
    \begin{equation}
        \hat{y}^{(0)} = \mathcal{F}\{y\}
    \end{equation}

    \item At each scale $s = 1, \dots, S$, bandpass filters $B_k^{(s)}$ extract orientation-selective responses for $k = 1, \dots, K$ directions, while a lowpass filter $L^{(s)}$ generates the coarser scale:
    \begin{equation}
        \hat{z}_k^{(s)} = B_k^{(s)} \cdot \hat{y}^{(s-1)}, \qquad \hat{y}^{(s)} = L^{(s)} \cdot \hat{y}^{(s-1)}
    \end{equation}

    \item Inverse Fourier transform reconstructs spatial responses:
    \begin{equation}
        z_k^{(s)} = \mathcal{F}^{-1}\{\hat{z}_k^{(s)}\}
    \end{equation}

    \item After the pyramid decomposition, both the high-pass (finest scale) and low-pass (coarsest scale) bands are removed. The remaining complex-valued orientation responses are converted to magnitude energy maps:
    \begin{equation}
        e_k^{(s)} = |z_k^{(s)}|^2 = \Re(z_k^{(s)})^2 + \Im(z_k^{(s)})^2
    \end{equation}

    \item At the last step each energy map $e_k^{(s)}$ is flattened and concatenated into a single feature vector:
    \begin{equation}
        \mathbf{v} = \texttt{concat}\left( \texttt{vec}(e_k^{(s)}) \ \forall \ s, k \right)
    \end{equation}
\end{enumerate}

This representation, which is used in the trajectory straightness analysis, encodes multi-scale,~orientation-specific contrast energy,  reflecting the known response of V1 simple cells.

\bibliographystyle{ieeetr}  
\bibliography{refs}  

\end{document}